\documentclass[sigconf]{acmart}

\pdfoutput=1

\usepackage{nopageno}

\usepackage{epsfig,endnotes}
\usepackage{type1cm}	
\usepackage{graphicx}     
\usepackage{xspace}     
\usepackage{balance}     
\usepackage{booktabs}     
\usepackage[tableposition=top,font={small},skip=8pt]{caption}     
\usepackage{siunitx}	
\usepackage{epsfig}
\usepackage{todonotes} 
\usepackage{framed}
\usepackage{balance}
\usepackage{newfloat}
\usepackage{amsmath, amssymb, eurosym}
\usepackage{enumerate}
\usepackage[vlined,linesnumbered,ruled,noend]{algorithm2e}
\usepackage{lscape}
\usepackage{subfigure}
\usepackage{array}
\usepackage{pifont}
\usepackage{color, colortbl}
\usepackage{alltt}
\usepackage{listings}
\usepackage{balance}
\usepackage{multirow}
\usepackage{hyperref}
\usepackage{enumitem}

\newcommand{\hb}{HB\xspace}
\newcommand{\hblong}{Header Bidding\xspace}
\newcommand{\DP}{Demand Partner\xspace}
\newcommand{\DPs}{Demand Partners\xspace}
\newcommand{\cshb}{Client-Side \hb\/\xspace}
\newcommand{\sshb}{Server-Side \hb\/\xspace}
\newcommand{\toolname}{\emph{HBDetector}\xspace}
\newcommand{\datasetAuctionSize}{800k\xspace}
\newcommand{\etc}{etc.}
\newcommand{\eg}{e.g., }
\newcommand{\ie}{i.e., }
\newcommand{\etal}{et al. }

\newcommand{\alexaTop}{35,000\xspace}
\newcommand{\sitesWithHB}{5,000\xspace}
\newcommand{\detectionPerc}{14.28\%\xspace}

\definecolor{heraldGreen}{rgb}{0.0,0.4,0.0}
\hypersetup{
	bookmarks=false, 
	colorlinks=true,%
	citecolor=black,%
	filecolor=black,%
	linkcolor=black,%
	urlcolor=black,%
	pagebackref=true,%
}
\def\ttfntsize{9}
\makeatletter
\let\oldtexttt\texttt
\let\texttt\@undefined
\makeatother
\newcommand{\texttt}[1]{\fontsize{\ttfntsize}{\ttfntsize}\oldtexttt{#1}}
\makeatletter
\let\oldtt\tt
\let\tt\@undefined
\makeatother
\newcommand{\tt}{\fontsize{\ttfntsize}{\ttfntsize}\oldtt}

\renewcommand{\shortauthors}{Pachilakis et al.}

\title[Measuring the Header Bidding Ad-Ecosystem]{No More Chasing Waterfalls: A Measurement Study of the Header Bidding Ad-Ecosystem}


\begin{document}	
\author{Michalis Pachilakis}
\affiliation{
	\institution{University of Crete / FORTH, Greece}
}
\email{mipach@ics.forth.gr}

\author{Panagiotis Papadopoulos}
\affiliation{
	\institution{Brave Software}
}
\email{panpap@brave.com}

\author{Evangelos P. Markatos}
\affiliation{
	\institution{University of Crete / FORTH, Greece}
}
\email{markatos@ics.forth.gr}

\author{Nicolas Kourtellis}
\affiliation{
	\institution{Telefonica Research, Spain}
}
\email{nicolas.kourtellis@telefonica.com}


\begin{abstract}
In recent years, \hblong (\hb) has gained popularity among web publishers, challenging the status quo in the ad ecosystem.
Contrary to the traditional waterfall standard, \hb aims to give back to publishers control of their ad inventory, increase transparency, fairness and competition among advertisers, resulting in higher ad-slot prices.
Although promising, little is known about how this ad protocol works:
What are \hb's possible implementations, who are the major players, and what is its network and UX overhead?

To address these questions, we design and implement \toolname: a novel methodology to detect \hb auctions on a website at real-time.
By crawling \alexaTop top Alexa websites, we collect and analyze a dataset of \datasetAuctionSize auctions.
We find that:
(i) \detectionPerc of top websites utilize \hb.
(ii) Publishers prefer to collaborate with a few \DPs who also dominate the waterfall market.
(iii) \hb latency can be significantly higher (up to 3$\times$ in median case) than waterfall.
\end{abstract}

\maketitle

\section{Introduction}

The largest portion of the digital advertisements we receive today on the Web follows a \emph{programmatic ad-purchase} model.
Upon a website visit, a real time auction gets triggered, usually via the real-time bidding (RTB) protocol~\cite{rtb}, for each and every available ad-slot on the user's display.
These auctions are hosted in remote marketplace platforms called Ad Exchanges (ADXs) 
that collect the bids from their affiliated Demand Site Platforms (DSPs). 
The highest bidder wins, and delivers its impression to the user's display.

However, there are more than one ad networks that can provide bids for an ad-slot.
In the traditional standard for ad-buying, called \emph{waterfalling}, the different ad networks (\eg ADXs with their affiliated DSPs) are prioritized in hierarchical levels ~\cite{waterfalling}.
Thus, when there is no bid from ad network \#1, a new auction is triggered for ad network \#2, and so forth.
Of course, apart from the auction-based ad purchase, there are still other non-programmatic channels like \emph{direct orders} from advertisers who run static campaigns for a certain number of impressions~\cite{static_ads}.
Through these channels, advertisers target not a user but the entire audience of a specific website (\eg an ad regarding Super Bowl on {\tt espn.com}).
Alternatively, if there is neither a direct order nor a bid in these auctions, the ad-slot may be filled via another channel for remnant inventory called \emph{fallback or backfill} (\eg Google AdSense)~\cite{backfill_ad}. 

The process of ad prioritization among the above different channels and ad networks in waterfall is managed through the publisher's ad server or Supply Side Platform (SSP) (\eg DoubleClick for Publishers (DFP)).
Priorities are typically set not at real time but based on the average price of the past purchases for each channel.
As a consequence, in waterfall not all ad partners have the ability to compete simultaneously.
Therefore, \emph{the publishers do not get the optimal charge price}, since an ad-slot may not be sold at the highest price (\eg if the winning bid in the auction of ad network \#1 is 0.2\$, the ad-slot will be sold even if there was a bid of 0.5\$ in ad network \#2).
Apart from the potential loss of revenue for the publishers, there is also a \emph{significant lack of transparency}.
Except from the winning bidder, the publishers do not know who else placed a bid for their ad-slot and for how much.
In addition, the lack of control restricts the publishers from choosing \DPs, or different sale channels in real time (\eg to get a high price through RTB when the quota of direct ads sold has not yet been depleted).

\begin{figure}[t]
	\centering
	\hspace{-0.3cm}
	\includegraphics[width=1.07\linewidth]{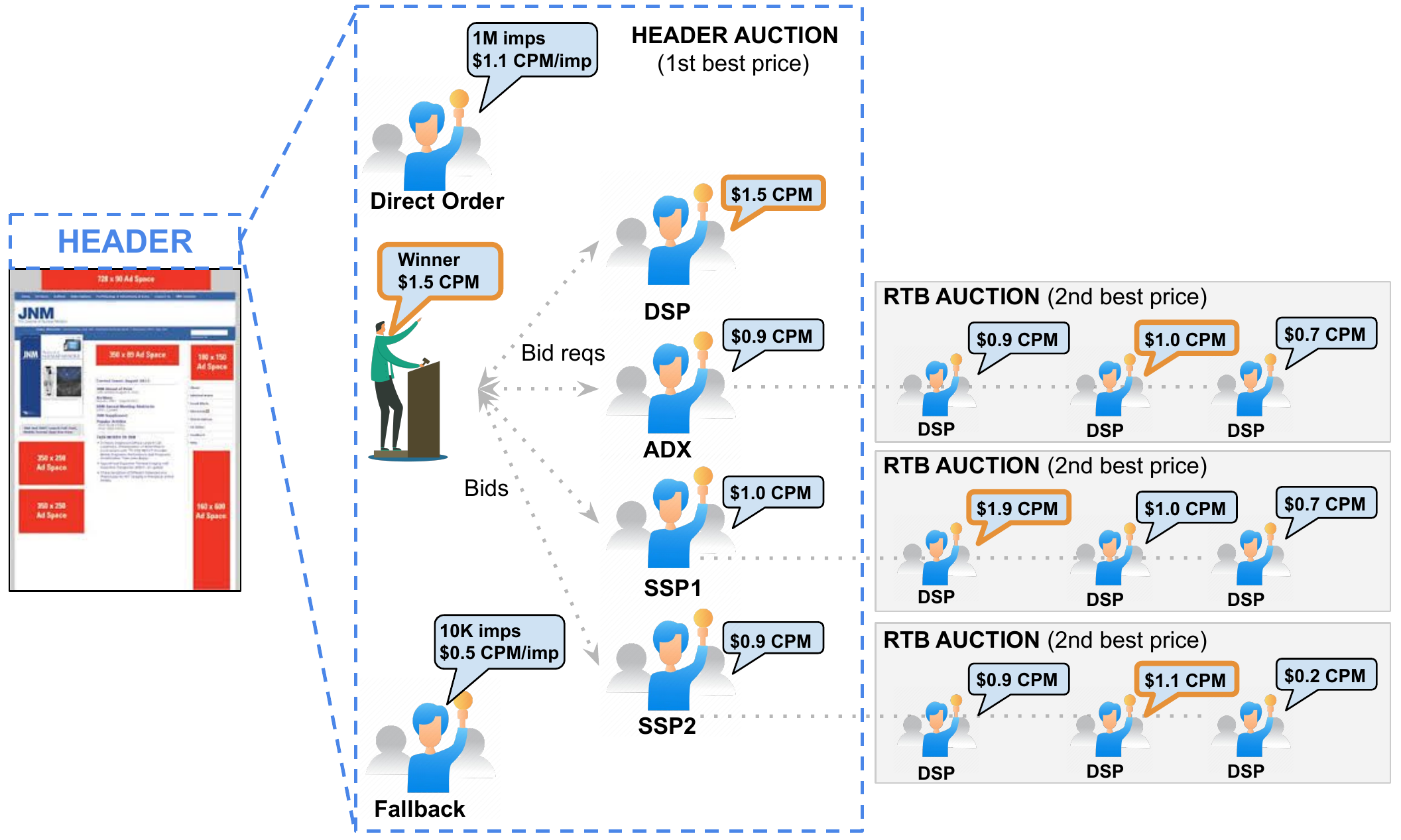}
	\caption{High level overview of the \hb. The absence of priorities aims to provide (i) fairness and higher competition among advertisers and (ii) increased revenue for the publishers.}
	\label{fig:hbidding}
\end{figure}

To remedy all the above, \emph{\hblong}~\cite{hb-understanding} (or \emph{parallel bidding} in mobile apps~\cite{parallelbidding}) has been recently proposed and has started to gain wide acceptance among publishers~\cite{hb-growth,hb-growth2,hb-growth3,hb-growth4}.
As depicted in Figure~\ref{fig:hbidding}, \hb is a different auction that takes place not on the ad server as in waterfall, but inside the header field of a HTML page, before anything else is loaded on the page.
It allows a publisher to simultaneously get bids from all sale channels (\eg direct orders, programmatic auctions, fallback) and \DPs (\eg DSPs, ADXs, ad agencies).
\hb not only gives the control back to the publisher but also allows higher revenues than waterfall, since it guarantees that the impressions with the higher price will be bought and rendered~\cite{hb-moreRev}.
On the advertiser's side, \hb promotes fairness since there are no priorities.
Consequently, any advertiser could win any auction, as long as it bids higher than others.
\hb enables small advertisers to also be competitive, compared to big advertisers who would have higher priority on the waterfall model.

Although there is a lot of research regarding the waterfall standard~\cite{imcRTB2017,lukasz2014selling-privacy-auction,DiffusionofUserTrackingDataintheOnlineAdvertisingEcosystem,olejnikbid,bashirtracing}, we know very little about the innovative and rapidly growing alternative of \hb.
How is it implemented?
What is the current adoption of \hb on the Web?
What is the performance overhead and how it affects the page rendering time? 
How many bids the average publisher can receive?
What are the average charge prices and how do these compare to the ones of the waterfall standard?
Which are the big players and how is the market share divided?

To respond to all these questions, we study the different existing implementations of \hb and we design \toolname: a novel methodology to detect \hb auctions on the Web.
Our approach aims to increase transparency on the ad-ecosystem, by exposing at real-time the internals of the new and rapidly growing \hb ad protocol: in which sites it exists, the prices and partners involved, etc.
Using \toolname, we crawl a number of popular websites, we collect a rich dataset of \hb-enabled websites.
Our tool helps us detect particular browser events triggered by the \hb libraries embedded in such webpages, along with the ad partners participating in the \hb and metadata for the auctions executed on these websites.
We analyze and present the first full-scale study of \hb aiming to shed light on how this innovative technology works and investigate the trade-off between the overhead it imposes on the user experience and the average earnings it brings to publishers. 
In this paper, we make the following main contributions:

\begin{enumerate}[leftmargin=0.5cm]
	\item
We propose and implement \toolname, the first of its kind Web transparency tool, capable of detecting \hb activity at real-time, on the Web.
We provide it as an open-sourced browser extension for Google Chrome\footnote{\url{https://www.github.com/mipach/HBDetector}}
.

\item 
	By running \toolname across \alexaTop top Alexa websites, we collect a dataset of \datasetAuctionSize auctions. 
This work is the first to analyze the adoption and characteristics of \hb.

\item 
	We extract a set of lessons on \hb:
(i) There are 3 different implementations of \hb: Client-side, Server-side and Hybrid \hb.
(ii) There is at least \detectionPerc of top websites that use \hb.
(iii) Publishers tend to collaborate with a small number of \DPs, which are already reputable in the waterfall standard.
(iv) \hb latency can be significantly higher (up to 3$\times$ in the median case, and up to 15$\times$ in 10\% of cases) than waterfall.

\end{enumerate}

\section{Background on \hblong}

In this section, we cover background knowledge required for our study regarding the most important aspects of \hb.

\begin{figure}[t]
	\centering
	\hspace{-0.3cm}
	\includegraphics[width=1.0\linewidth]{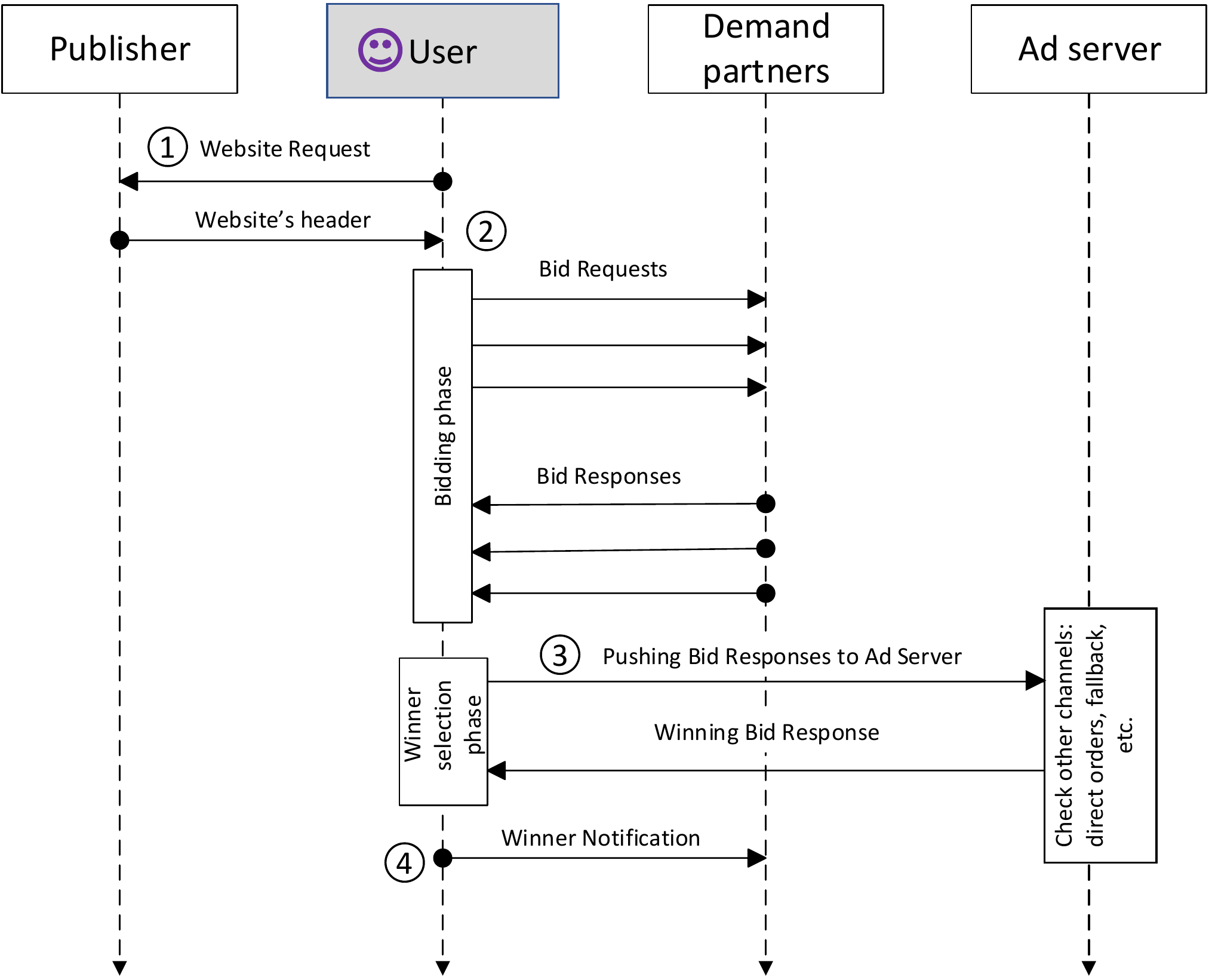}
	\caption{Flow chart of the \hblong protocol.}
	\label{fig:hb_UML}
\end{figure}

\subsection{\hb Protocol Description}
Contrary to the traditional waterfall standard, in \hb the ad auction does not take place in a remote ADX, but on the user's browser.
The \hb process, depicted in Figure~\ref{fig:hb_UML}, is the following:

\textbf{Step 1}: When a user visits a website, the HTML page is fetched.
As soon as the header of the HTML is rendered in the browser, user tracking code and the third-party library responsible for the procedure of the \hb is loaded as well.

\textbf{Step 2:} Then, the \hb library sends (in parallel) HTTP POST requests to the \DPs (\eg DSPs, ad agencies, ADXs which conduct their own RTB auctions) requesting for bids.
These bid requests also include information about the current user (such as interests and cookies).
Such information can be used by the \DPs to decide whether and how much they will bid for an ad-slot in the particular user's display.
Note, that if a \DP does not respond within a predefined time threshold, its bid is considered late and not taken into account. 

\textbf{Step 3:} As soon as the \DPs respond with their bids (and their impressions), the collected responses are sent to the publisher's ad server.
The ad server will check the received bids and compare with the floor price agreed with the publisher, to decide if the received prices are high enough~\cite{ratko17}.
If the floor price is met, the \hb process was successful and the ad-slot is satisfied.
Alternatively, the ad server can check the rest of the programmatic (or not) available channels (\eg direct order, RTB, fallback) and will find the best next option for the specific ad-slot.
This step entails communicating with SSPs for available direct orders which can provide higher revenues to the publisher than regular RTB auctions.
The ad server can also communicate with \DPs for RTB auctions, or other SSPs who can provide fallback ads, such as Google AdSense, or even house ads.

\textbf{Step 4:} As soon as the impression is rendered on the user's display, a callback HTTP request notifies the winning \DP that its impression was rendered successfully on the user's browser, and the ad price that was charged (winner notification).

In theory, with this new protocol, the publisher has total control over the ad inventory they provide, knowing exactly how much the \DPs value each slot, and the actual amount of money they are willing to pay for it.
In addition, there is full transparency, since the publisher can have access to all bids and decide at real time the best strategy it should follow without the need to trust any intermediaries.
In the future, \hb could provide the means to publishers to reduce advertising that is not suitable for, or does not match the semantics of their websites, and even curb malvertising.
However, as we will show later in Section~\ref{sec:types-of-hb}, this transparency and control is not always applicable under the various types of \hb we have detected.

\subsection{\hb Implementation \& Performance}

To implement the above protocol, publishers need to include \hb third-party libraries in their webpages.
Although there does not exist a common standard for \hb yet, the great majority of publishers use the open-source library of {\tt Prebid.js}~\cite{prebid}, supported by all major ad companies.
This library includes:
(i) The core component which is responsible to issue the bid requests and collect the responses, which are later sent to the publisher's ad server.
(ii) The adapters which are plugged into the core and provide all necessary functionality required for each specific \DP. 
{\tt Prebid.js} is supported by more than 200 \DPs  (\eg AppNexus, Criteo, OpenX, PulsePoint) that provide their own adapters~\cite{adapters}.

We note that in traditional waterfall, the auction information is opaque to the client and the only information that can be inferred (if at all) is through the parameters of the notification URL (which acts as a callback to the winning bidder).
In contrast, in the \hb, and due to bidder responses, browser DOM events are triggered that contain metadata directly available at the user browser, and can be used to clearly distinguish between waterfall and \hb activity.

The non-hierarchical model of \hb produces much more network traffic than the waterfall standard. 
Indeed, \hb sends one request for each and every collaborating \DP.
This can result to an increased page latency, especially when some \DPs take too long to respond. 
To make matters worse, as soon as they receive a bid request, some of these \DPs may run their own auctions inside their ad network, with their own affiliated bidders (as depicted in Figure~\ref{fig:hbidding}). 
This increased page latency raises significant concerns.
Indeed, 40\% of the publishers already mention that such latency is capable of impacting their users' browsing experience~\cite{latencyHB,latencyHB2,latency3}.

It is worth noting, that \hb technology is still in its early stages and many ad networks are technologically not ready to move completely from the waterfall model to participate in this new model.
In order not to miss bids from such networks, some ad mediators (e.g., Appodeal) mix the two techniques in an attempt to provide waterfall compatibility during this transitional period~\cite{parallelbidding}.

\section{Methodology for Measuring \hb}\label{sec:methodology}

In this section, we outline our methodology for detecting \hb on webpages, and our effort to crawl top Alexa websites for \hb activity.

\subsection{Detection Mechanism}\label{sec:detection}

In order to detect if a webpage is using \hb for delivering ads to its users, we need to detect \hb-related activity originating from the said webpage.
As explained above, the \hb activity is performed over different channels than ad protocols such as RTB, using a library (implemented in JavaScript) embedded in the header of the page.
Therefore, by monitoring the events triggered by such libraries, we can confidently distinguish \hb activity from other models such as waterfalling.

There are three main ways to detect if \hb is present in a webpage:
\begin{enumerate}
	\item Perform static analysis of the page and identify tags of scripts that load known \hb libraries.
	\item Detect DOM-related events that are triggered due to \hb embedded in the webpage.
	\item Detect web requests sent from the page to \hb entities.
\end{enumerate}

The first method is straightforward to implement with the following steps: Download the webpage source code and use regular expressions to detect all known \hb libraries.
However, we note that just detecting these libraries is not enough, as false positives or false negatives could occur.
For example, static analysis is prone to false positives such as non \hb libraries being misnamed using \hb-related names, or \hb-related libraries appearing in the HTML code but not executed 
Similarly, static analysis is vulnerable to false negatives such as renamed \hb libraries to names that are not known yet, or new \hb libraries that do not match our \hb-related keywords from known libraries.
To avoid such potential false positives and negatives, we chose not to use static analysis in the \toolname.

The second method is more difficult to implement, but offers better detection rates with reduced false positives and negatives, and thus, harder to evade.
This method monitors the DOM events that are triggered in a webpage, events that are sent to notify the code of interesting activity that has taken place on the page.
Events can represent everything from basic user interactions to automated notifications happening on the page.
Most \hb libraries trigger events in several phases of an auction (initiation of the auction, bid collection, winning bidder, etc.).
If such an event is detected, we are certain that it is because of \hb.
Even better,  by ``tapping'' on these events~\cite{tapping}, we can collect information about \hb that the first method is not able to detect.

The third method is similar to the second, but operates at a different level in the browser: monitor the web requests of a page in real-time, and detect all the request sent to and received from known \hb \DPs.
By constructing a list containing all the known \DPs, we can check all the incoming and outgoing WebRequests to the browser, and keep the relevant to \hb.

\begin{figure}[t]
	\centering
	\hspace{-0.3cm}
	\includegraphics[width=.8\linewidth]{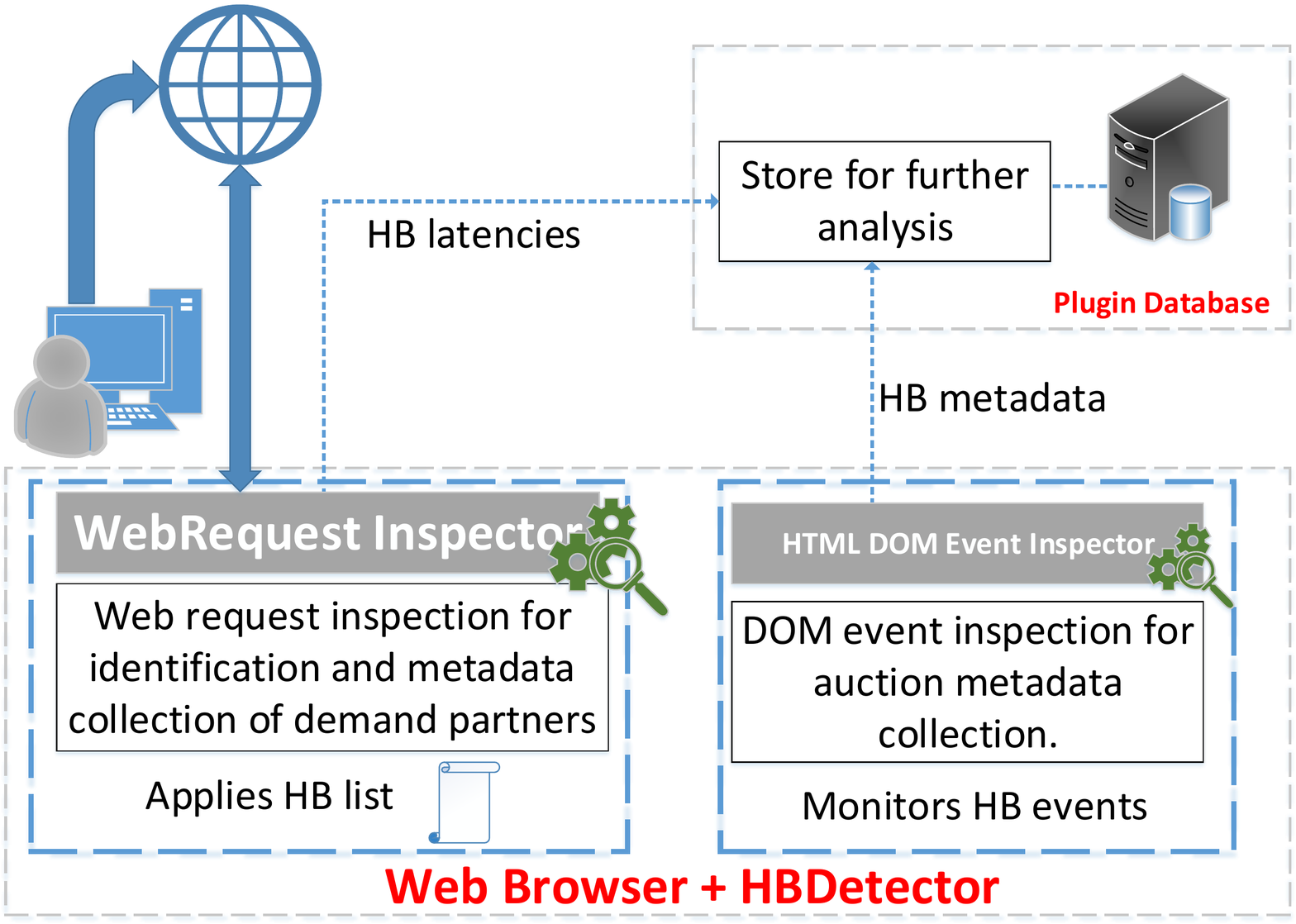}\vspace{-0.3cm}
	\caption{Overview of the \toolname mechanism.
	After the user accesses a webpage, all the incoming and outgoing WebRequests are inspected to detect \hb partners.
	A content script is also injected in the header of the webpage to detect \hb events about the auction performed.}
	\label{fig:detection_mechanism}
\end{figure}

In this paper, we implemented \toolname, a tool which combines the 2nd and 3rd methods to increase detection performance.
An overview of the tool is illustrated in Figure~\ref{fig:detection_mechanism}.
\toolname adds a content script in the header of each webpage when the page is loaded.
This script monitors the webpage's activity for various events and requests sent and received by the page, keeping the ones relevant to HB (e.g., incoming responses from DSPs for HB auctions).
When such DOM events are triggered (which is a first sign of HB activity), the tool filters the web requests triggering these events by checking the parameters included in them.
HB libraries use predefined parameters such as ``$bidder$'', ``$hb\_partner$'', ``$hb\_price$'', etc., which are not used by other ad-protocols such as RTB.
Thus, the tool keeps all web requests that triggered a DOM event of a \hb library, and also include \hb parameters.
It then proceeds to extract the values from these parameters for analysis.
These parameters are typically fixed for each HB library, and all HB partners must use them as such, to participate successfully in HB auctions with that library.
In contrast, in the RTB protocol, the parameter names used in the notification URLs are DSP dependent and do not utilize DOM events.

From the available \hb libraries, we examined {\tt prebid.js} (and its variants), being the most famous one (64\% of client-side wrappers are built on prebid~\cite{industry_adzerk}), as well as {\tt gpt.js} and {\tt pubfood.js} libraries for their available codebase and/or documentation.
We decided to focus our more in-depth reverse-engineering on {\tt prebid.js} due to its popularity, available documentation and open-source code and APIs~\cite{gpt, prebidapi}.
By performing code and documentation analysis for the \hb libraries that have such material available, we identified the following list of \hb events that our tool can detect:
\begin{itemize}
	\item \textit{auctionInit}: the auction has started
	\item \textit{requestBids}: bids have been requested
	\item \textit{bidRequested}: a bid was requested from specific partner
	\item \textit{bidResponse}: a response has arrived
	\item \textit{auctionEnd}: the auction has ended
	\item \textit{bidWon}: a bid has won
	\item \textit{slotRenderEnded}: the ad's code is injected into a slot
	\item \textit{adRenderFailed}: an ad failed to render
\end{itemize}

In this work, we focus on three of these events: \textit{auctionEnd}, \textit{bidWon}, and \textit{slotRenderEnded}.
The \textit{auctionEnd}, as its name states, is triggered after the auctions for the ad-slots have finished, i.e., the \DPs have submitted their offers.
The \textit{bidWon} event is triggered after the winning \DP has been determined.
Finally, the \textit{slotRenderEnded} event is triggered when an ad has finished rendering successfully on an ad-slot.
By analyzing these events, which can only be triggered by \hb activity and not other libraries, we were able to collect several metadata about the auctions, such as the \DPs who bided, the ones who won, the CPM (cost per million impressions in USD) spent, the ad size, currency, dimensions, \etc

\begin{table}[]
	\caption{Summary of collected data by crawling top Alexa webpages using the \toolname for \hb-related activity.} \label{tab:data-summary} \vspace{-0.2cm}
	\begin{tabular}{ll}
		\toprule
		Data & Volume \\ \hline	
		\rowcolor[HTML]{EFEFEF} 
		\# of websites crawled	&	\alexaTop 	\\
		\# of websites with HB	&	4998		\\
		\rowcolor[HTML]{EFEFEF}
		\# of auctions detected	&	798,629		\\
		\# of bids detected		&	241,392	\\
		\rowcolor[HTML]{EFEFEF}
        \# of competing \DPs 	&	84		\\
		\# weeks of crawling		&	5		\\
		\bottomrule
	\end{tabular}
\end{table}

We also constructed a list with all the known \hb \DPs.
We collected and combined several lists used by \hb tools designed to help publishers fine tune their \hb on their websites.
Using this list, we can infer all the \textit{WebRequests} about \hb without altering them, in order to detect when a request to a \DP is sent, 
and when an answer is received.
The \toolname is written in a few hundred lines of JavaScript as a Google Chrome browser extension.

\textbf{\toolname limitations:}
The tool does not analyze all libraries used by the \hb ecosystem due to unavailability of documentation and/or code.
Also, it cannot capture new DOM events if they get added to existing libraries it is analyzing.
Finally, it cannot capture current DOM events if the events change format or parameters they are using.
In addition, the tool does not capture waterfalling RTB activity, and therefore, does not allow direct comparison of the two protocols with respect to \DPs involved, ad-prices, etc.
We plan to address these limitations in a future version of the tool.

\subsection{Data Crawling}

We used our tool to detect which websites employ \hb, by crawling a set of websites, based on a large top list purchased from Alexa~\cite{alexa_rank} on 01/2017.
Given the changes anticipated in such website ranking list and especially in its long tail~\cite{scheitle18toplists}, we focus on the head of the Alexa list, to capture a more stable part of the ranking distribution through time.
Due to equipment, network and time costs, we limited this list to \alexaTop domains to crawl per day, during Feb'19.
To confirm the representativeness of this older list, we compared it with the top 35k domains in 2017-2019 from~\cite{scheitle18toplists}, and found that it has an overlap of 78.36\%(06/2017), 62.10\%(06/2018), 58.36\%(02/2019) and 55.34\%(06/2019).

We used selenium and chromedriver loaded with \toolname in order to automate the crawling.
We initiated a clean slate instance before visiting each website, in order to keep the crawling process stateless (no previous history, no cookies, no user profile).
When a webpage is visited, the crawler waits for the page to be completely loaded, and then allows an extra five seconds, in case additional content needs to be downloaded or pending responses to be concluded.
We set the page load timeout to 60 seconds, so that if the page is not fully loaded in one minute, the crawler proceeds to the next webpage in the list, after killing the previous instance and initiating a new, clean instance.

With this crawling process, we detected \hb in $\sim$\sitesWithHB (14.28\%) of the websites, in a well-distributed fashion.
In particular, \hb was found in 20-23\% of the top 5k websites, 12-17\% for the top 5k-15k, and 10-12\% for the rest.
Indeed, new top websites not included in this 35k list may have already adopted HB, leading to an underestimation of today's adoption rate.
However, as found in our results in Sec.~\ref{sec:hbadoption} were we use top 1k Alexa lists for 6 years, we show similar \hb adoption rate with the head of the top 35k list, giving credence to our results.
Then, we crawled these 5k websites every day for a period of 34 days in Feb'19, collecting metadata about the \hb auctions, and performance exhibited from the various websites using \hb.
In Table~\ref{tab:data-summary}, we provide a summary of the data collected.

We note that we detected \datasetAuctionSize auctions but received 241k bids.
One could expect that each auction should have at least a bid.
Indeed this would be the case if actual users were involved and \DPs were interested in them.
However, there are cases where bidders may avoid bidding when they know nothing about the user.
In our case, we are interested in the vanilla case using a clean state crawler and no real user profiles.

\section{The 3 Facets of \hblong}\label{sec:types-of-hb}

In this section, we analyze the crawled data and present results and observations we have made about the \hb adoption over time and types of \hb we identified from our exploration.

\subsection{\hblong Adoption}\label{sec:hbadoption}

\begin{figure}[t]
	\centering
	\includegraphics[width=.7\linewidth]{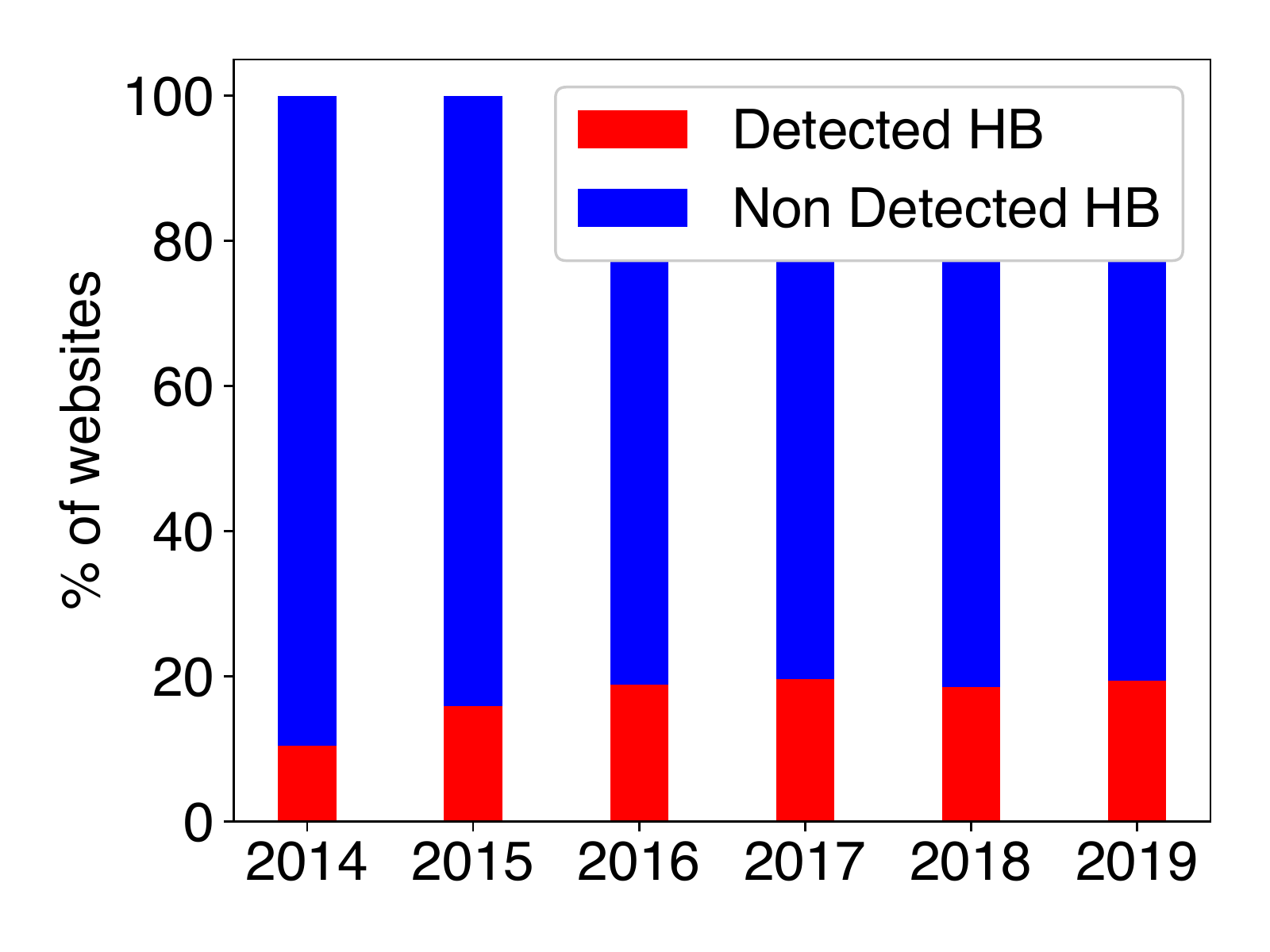}
	\vspace{-0.3cm}
	\caption{\hblong adoption in the last six years for the top 1k Alexa websites of every year.}
\label{fig:HBadoptability}
\end{figure}

Since this is a new programmatic ad-protocol (standardized in 2014\cite{HBhistory}), we explore the general adoption of \hb through the last 6 years.
To do that, we downloaded snapshots of selected lists of webpages using the \textit{Wayback Machine}~\cite{wayback}. 
Due to the involved network and time cost to crawl from the Wayback Machine, we focused on the top 1,000 publishers based on Alexa rankings, made available in a recent study~\cite{scheitle18toplists} and {\url{https://toplists.github.io/}}.
The list of top publishers was selected on a fixed day per year (6/6/2019, 6/6/2018, \etc).
Since these historical webpages were static, we performed a static analysis looking for \hb libraries and components in their websites' code.
Someone could also try an analysis using the \toolname, by attempting to render each website, or even fingerprinting the libraries.
However, such analyses: i) Take more time to execute than static analysis. ii) The webpage must be renderable and its components must work (scripts should be downloadable, scripts should not fail to run, the page should not call unresponsive servers, etc.).
Therefore, dynamic analysis cannot be applied on historical pages ``played back'', with potentially deprecated libraries or other scripts embedded, third-party partners not responding, etc., and expect 100\% correctness on the results collected.

Figure~\ref{fig:HBadoptability} shows the yearly breakdown of \hb found in these websites.
Interestingly, we observe a steady increase of the \hb adoption.
About 10\% of these websites were early adopters and started using \hb 6 years ago.
After the breakthrough of 2016, when \hb\ became popular\cite{HBbreakthrough}, there is a steady 20\% of the websites using this ad protocol.
These adoption rates, and the general rate of 14.28\% in the 35k list, match industry-claimed numbers of $\sim$15\% in the last 15 months (14.66\% in Jan'18 - 15.84\% in March'19, computed for the top 1k out of 5k top Alexa websites that serve programmatic ads) for the US market~\cite{emarketer_industry, industry_adzerk}.

We note that \toolname catches 100\% of the \hb activities for the libraries analyzed.
Indeed, there are websites which could be using \hb libraries that we didn't analyze at the time of data collection, and therefore were not flagged as \hb-enabled websites.
This means we get 100\% precision but not 100\% recall.
However, the \hb adoption experiment using the 1k lists shows a rate that aligns with the overall \hb adoption rate in the 35k list, and these two rates closely match what industry is claiming.
These observations point to low false positive and negative rates, and that the data collected by \toolname (i.e., using dynamic analysis) have high recall rate and provide a representative picture of the \hb ecosystem at the time of each crawl.

\subsection{Types of \hblong Detected}

Our in-depth investigation of the \hb ecosystem and the data collected revealed that this new programmatic ad protocol is currently being deployed in three facets:
(i) \cshb, (ii) \sshb, and (iii) Hybrid \hb.
This finding matches the 3 types of \hb wrappers (client-side, server-side and hybrid) suggested by industry reports~\cite{goodwaygroup}.
In the \cshb and Hybrid \hb models, the ad auctions are transparent, so we can distinguish them with a high degree of certainty due to the events sent and received by the browser.
On the other hand, on \sshb model it is less clear, since most of the ad-related actions happen at the server.
However, after inspecting the responses received by the browser, we can discover the parameters referring to HB (e.g. hb\_partner, hb\_price, etc.).
Next, we analyze each facet, including the steps taken for the protocol's execution, and potential consequences it may have.

\subsection{\cshb}

\begin{figure}
	\includegraphics[width=1.05\columnwidth]{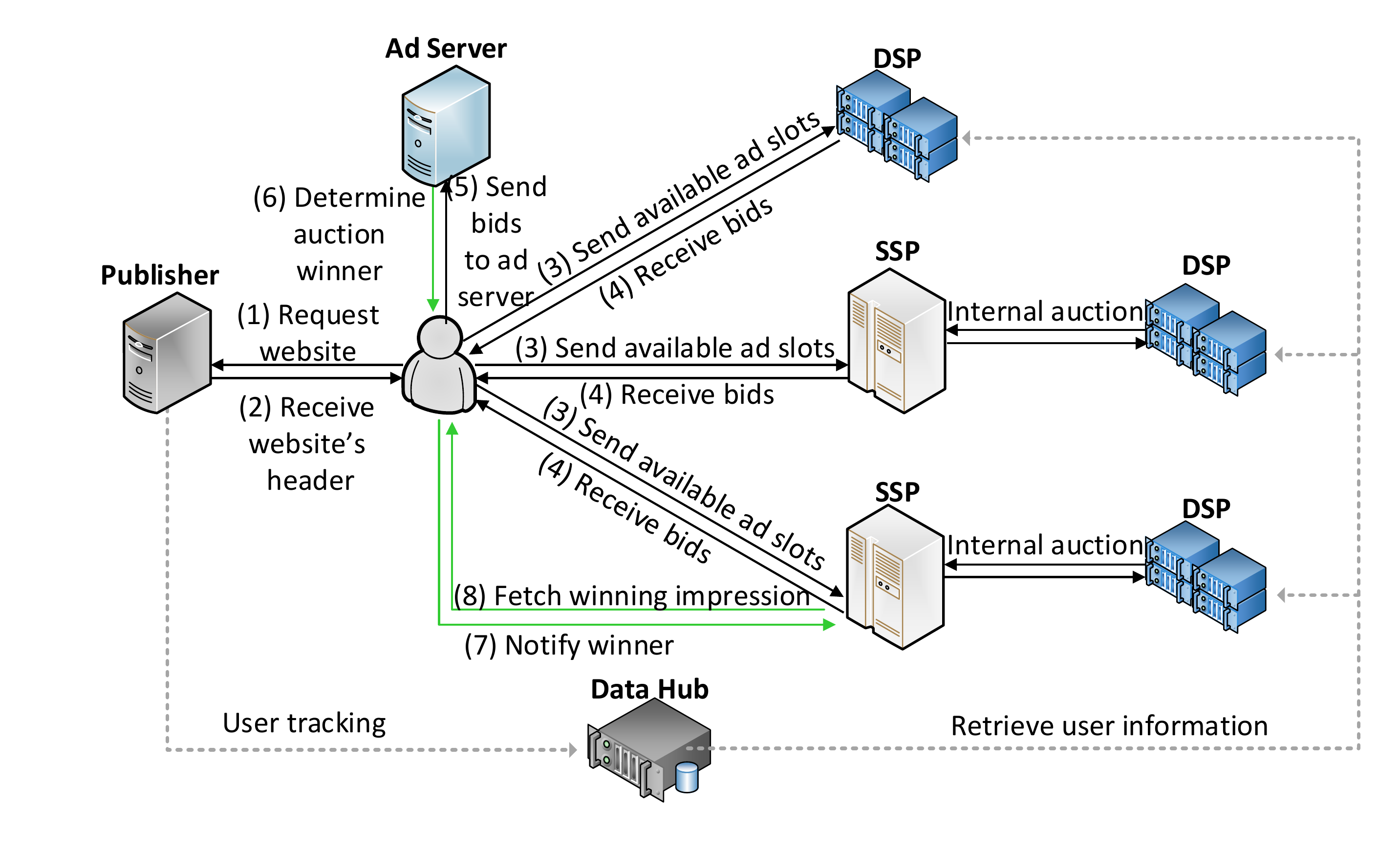}
	\vspace{-0.4cm}
	\caption{\cshb overview and steps followed.}
	\label{fig:csHB}
\end{figure}

In \cshb, as the name implies, the \hb process happens in the user's browser.
As illustrated in Figure~\ref{fig:csHB}, during this \hb type, the user's browser executes 8 steps, including the initiation of the \hb auction, receiving of bids from \DPs and notifying the winning partner.
\cshb's main goal is to improve fairness and transparency.
Publishers can choose the \DPs they want to collaborate with, regardless of their market cap.
What matters is if their bids are competitive enough.
Also, because the whole \hb process is performed at the client side, and then sent to the publisher's ad server, it is completely transparent to the publisher and, in theory, to the user.

The publisher can know at any time which partners bid, for which ad-slots they were interested, how much they were willing to pay, etc.
On the down side, \cshb is harder to set up.
Publishers need to have good technical understanding to set up and tune their \hb library.
Also, they need to operate their own ad server, a task which is not trivial.
Finally, because of the increased number of messages to be exchanged, or due to a bad configuration in the \hb library, longer latencies may be observed.

From the regular end-user's point of view, the only thing that can be observed is an increased latency for the loading of the webpage when it employs \cshb.
However, the regular user cannot be aware of all the \hb (and other ad-tech) activity happening in the background.
This is where our \toolname tool can help increase transparency of the protocol from the point of view of the end-user, and measure non-obvious aspects such as the communication and time overhead for the browser during \hb, winning bids, etc.

\subsection{\sshb}

\begin{figure}
	\centering
	\includegraphics[width=1.05\columnwidth]{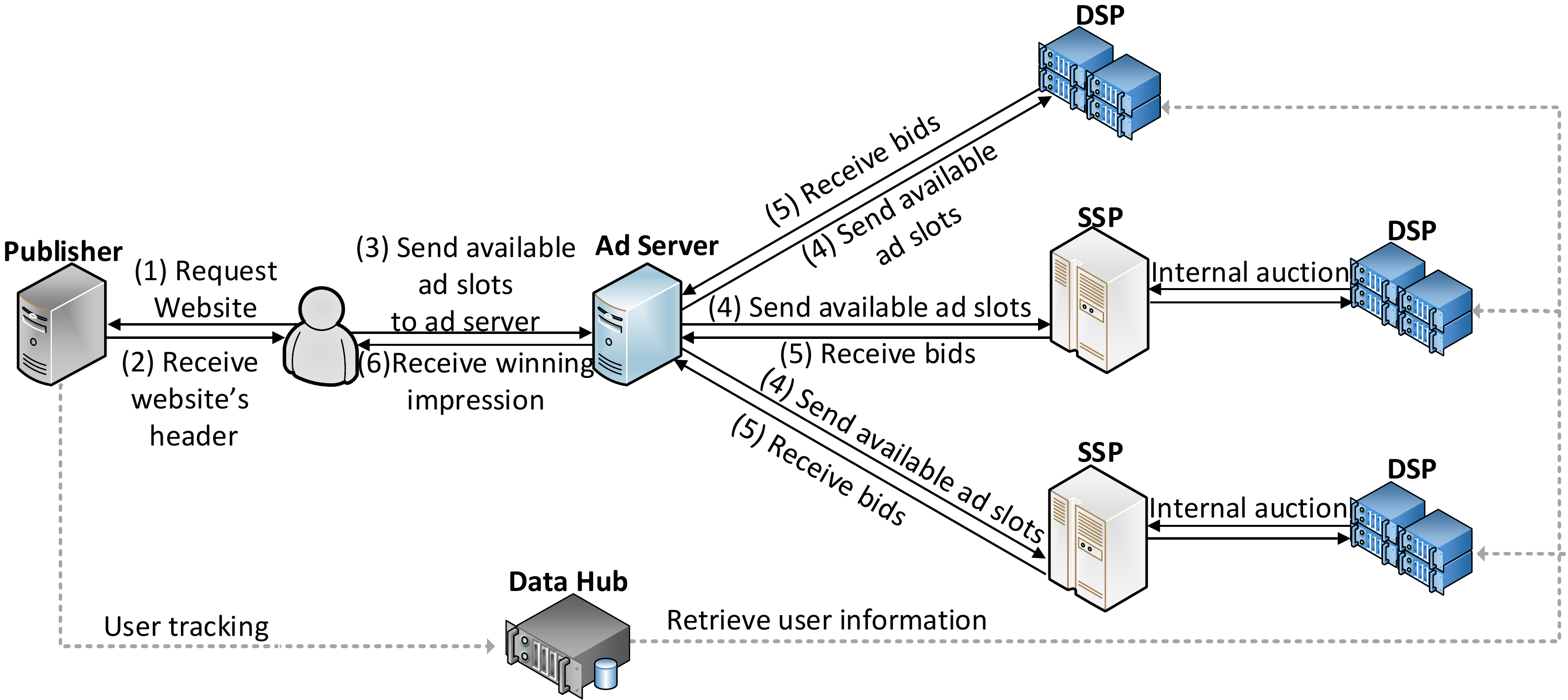}
	\vspace{-0.4cm}
	\caption{\sshb overview and steps followed.}
	\label{fig:ssHB}
\end{figure}

In \sshb, a single request is sent to a \DP's server, which is responsible to do the whole \hb process  and send back to the client only the winning impressions.
As \DPs, in this scenario, we consider all possible ad partners (SSPs, DSPs) that take part in the auction.
Figure~\ref{fig:ssHB} shows the \sshb model and the steps performed by the user's browser.
The careful reader will note a similarity of this model with \cshb with one \DP.
To distinguish \sshb from \cshb, we check the responses sent back from the \DP involved to the browser, to filter out bid responses (which would reveal \cshb cases).
This filtering using \hb-related keywords, also ensures that we are not mixing \hb with traditional waterfall activity.
Obviously, in this model the publisher needs to trust that the \DP (i.e., the server handling all requests) is honest, will not execute waterfall in the backend instead of \hb, and will select the best bids as winners, thus providing the best possible profits to the publisher.

\sshb requires the least effort from the publishers to setup their \hb.
However, in exchange for setup convenience, it reduces transparency to the minimum, since the publishers have no way of knowing the \DPs participating in the auctions or their actual bids.
Publishers don't need to tune their library, nor set up an ad server.
They just add to their webpage a pre-configured library, provided by the \DP they choose to collaborate with.
Also, this setup could make small players less competitive, compared to big ones with better infrastructure and higher influence to the market, because publishers could tend to trust the latter ones.
In effect, the \sshb has re-enabled the dominant players in RTB to regain control of the ad-bidding process which was momentarily transferred on the user browser.

From the end-user's point of view, this setup lacks transparency and does not offer many insights on how the whole \hb process either works, performs, or what impact it has on the user's browser: all auctions are done in the background, at the ad server's side.
This setup brings back the pros and cons of the typical RTB with ADXs playing the crucial and controlling role in the protocol.

\subsection{Hybrid \hb}

\begin{figure}
	\centering
	\includegraphics[width=1.05\columnwidth]{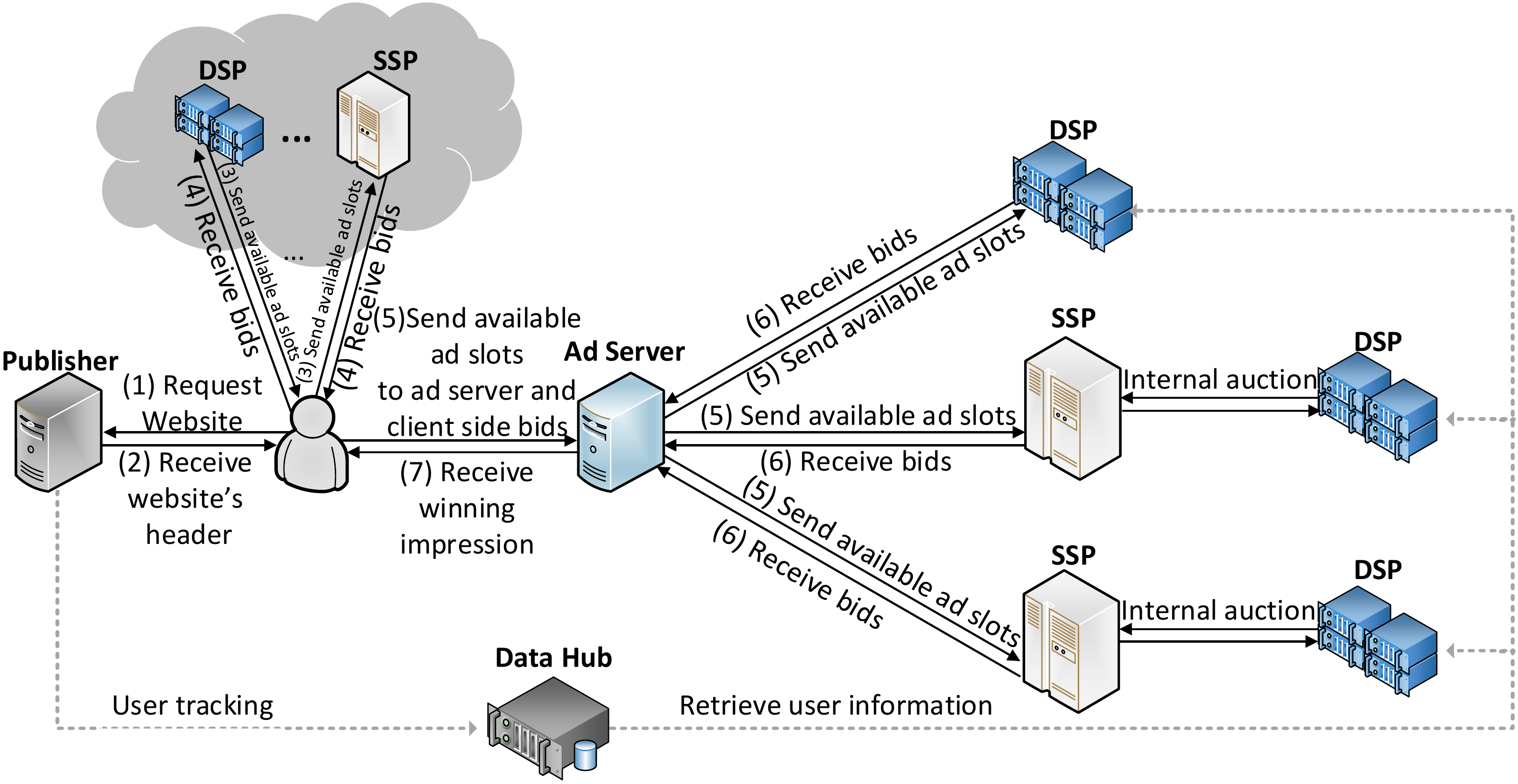}
		\vspace{-0.4cm}
	\caption{Hybrid \hb overview and steps followed.}
	\label{fig:hybridHB}
\end{figure}

\begin{figure*}[t]
	\centering
	\begin{minipage}{0.32\textwidth}
		\centering
		\includegraphics[width=1.05\linewidth]{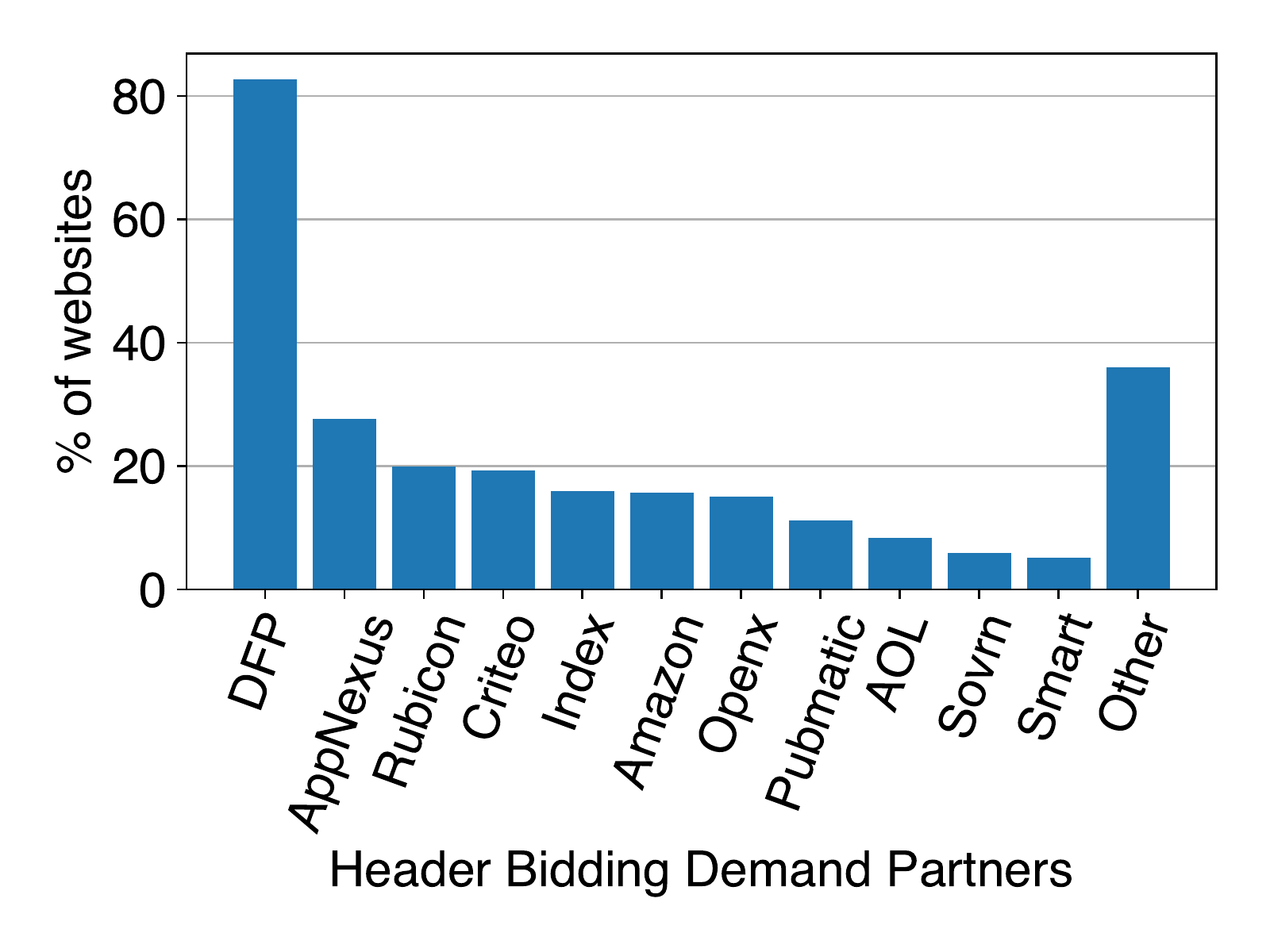}
		\caption{Top \DPs in \hb. Google (with DFP) is present in 80\% of websites with \hb.
			The rest of \DPs (N=73) have presence in  36\% of the websites with \hb.}
		\label{fig:topPrt}
	\end{minipage}
	\hfill
	\begin{minipage}{0.32\textwidth}
		\centering
		\includegraphics[width=1.05\linewidth]{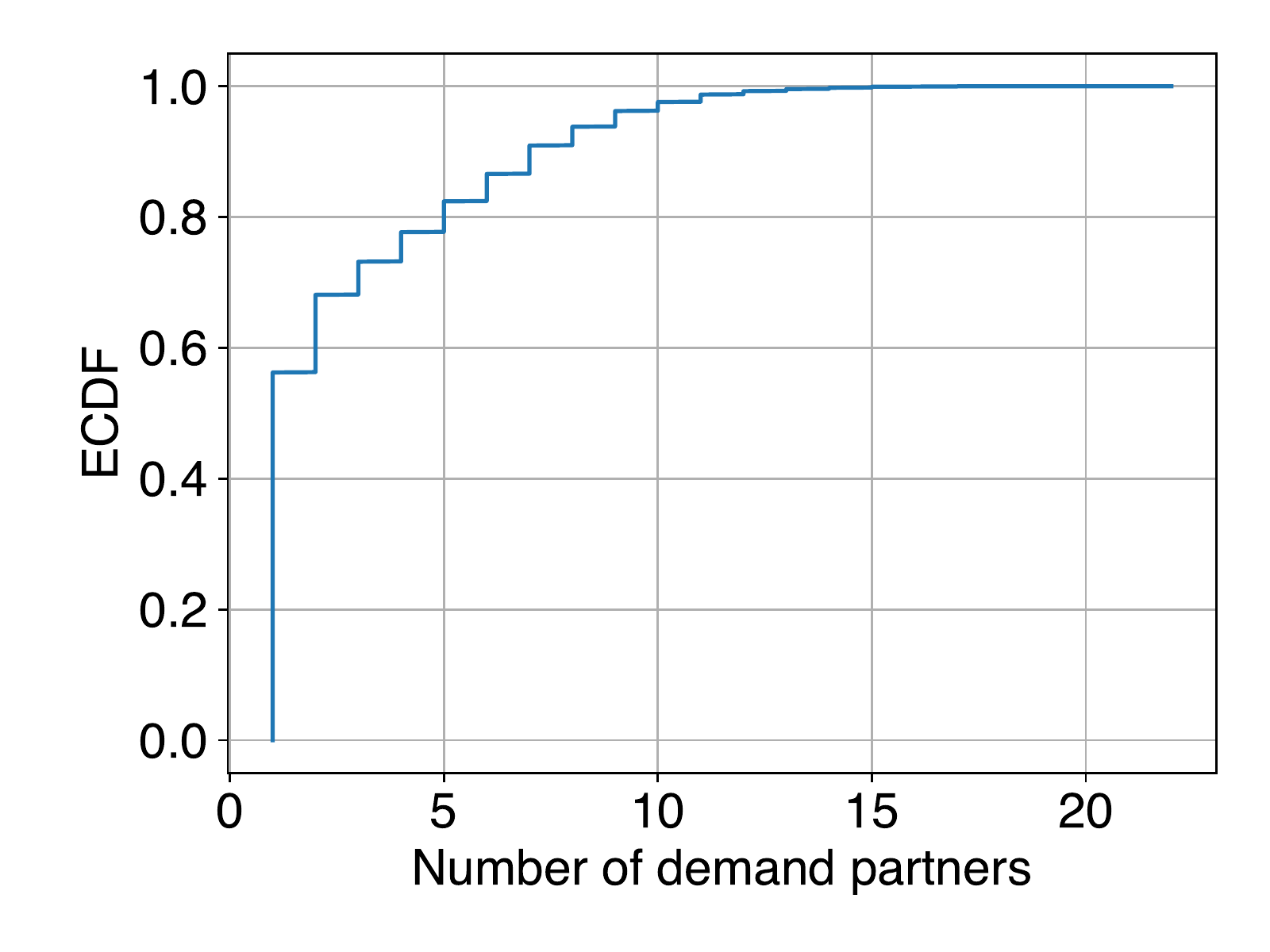}
		\caption{\DPs per website that employes \hb.
		More than 50\% of publishers use only one \DP, but some use as many as 20.}
		\label{fig:numberPrt}
	\end{minipage}
	\hfill
	\begin{minipage}{0.32\textwidth}
		\centering
		\vspace{0.2cm}
		\includegraphics[width=1.05\linewidth]{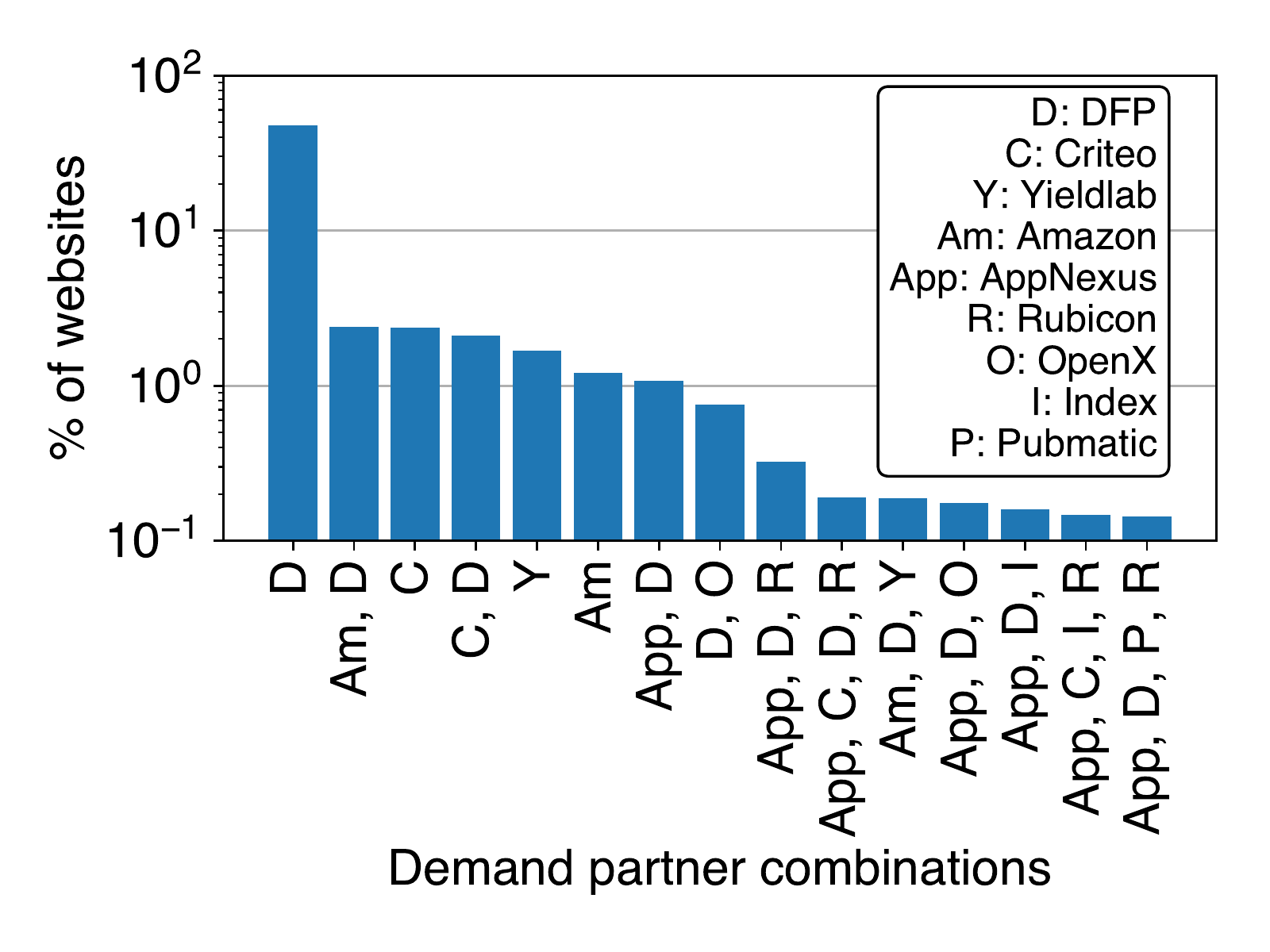}\vspace{-0.2cm}
		\caption{Most frequent \DPs combinations. DFP owns the majority of the market (48\%). Criteo follows with 2.37\% and Yieldlab with 1.68\%.}
		\label{fig:DPcombinations}
	\end{minipage}

\end{figure*}

As its name states, this is a hybrid model that combines \cshb\/ with \sshb\/ (Figure~\ref{fig:hybridHB}).
In this model, the user fetches the webpage which then requests bids from independent \DPs (as in the \cshb model).
When the browser (\hb library) receives the bid responses, it sends them to the ad server along with the available slots.
The ad server then performs its own auction (as in the \sshb model) and picks the final winning impression(s) from all collected bids (both from client and server side).
This model tries to combine the pros of \cshb and \sshb, while avoiding their cons.
It is a semi-transparent model with a certain degree of fairness, which requires a moderate degree of effort for the setup. 
Publishers can choose the \DPs they will collaborate with directly, so they can know the bids they are willing to pay.
Also they don't need to operate their own ad server, so the programmatic effort is reduced to tuning with the selected \DPs.

\subsection{Facet Breakdown}

The 3 facets of \hb that we observed and described above, have the following breakdown as detected from the \toolname in the wild (no other cases were observed that could comprise a 4th category).
We find that the \sshb currently comprises the larger portion of the market with 48\%.
Then, the Hybrid \hb is second with 34.7\%, and the \cshb is third with 17.3\%.
This means that publishers prefer the centralization and control offered by a server-side (or hybrid) model, which imposes a smaller overhead and increases speed of transactions.

Indeed, the actors that provide both HB and waterfalling options need to respect the protocols' guidelines, otherwise they won't participate successfully in the HB process.
Depending on the model they are called to use in each auction, they have to use the appropriate notification channel and parameters to notify the browser.
As we will see in the next section, this highly skewed breakdown towards server-side or hybrid is due to the presence of Google's DFP, which participates in many of these \hb auctions.

\section{Analyzing the \hb ecosystem}\label{sec:analysis}

Here, we analyze the data crawled in different dimensions:
\begin{itemize}
\item Number, diversity and combinations of \DPs participating in \hb (Section~\ref{sec:demand-partners})
\item Latencies measured with respect to overall \hb process, publishers and participating partners (Section~\ref{sec:latencies})
\item Auctions performed, bids received, bids taken into account or got lost (Section~\ref{sec:ad-auctions})
\item Properties of ads delivered: ad-slot prices paid and comparison with RTB prices. (Section~\ref{sec:ad-prices})
\end{itemize}

\subsection{\DPs Involved in \hb}\label{sec:demand-partners}

As a next step, we examine the properties of \DPs across the websites crawled and investigate who are the dominant \DPs, how many participate per website, and how they are combined together per webpage.

\noindent{\bf Who dominates the market?\\}
First, we examine the popularity of each \DP across all websites.
We define as popularity the percentage of sites that a given \DP participates in the site's \hb process.
In total, we find 84 unique \DPs.
Figure~\ref{fig:topPrt} shows the 11 most popular \DPs. As we can see, 
Google's DoubleClick for Publishers (DFP) is the most popular partner, with more than 80\% of publishers utilizing it.
The DFP can be used both as an ad server and as a server-side \hb solution.
Thus, it is not strange that most of the publishers choose this option over setting their own ad server.
We can also see that the list of top \DPs is full of popular partners that can be found also in the waterfall standard, as presented in past works\footnote{AppNexus, Index, Amazon, Rubicon, OpenX, AOL, Criteo, Pubmatic, and Sovrn, which match exactly what the industry claims as the top \hb bidders in Aug'19 reports~\cite{industry_adzerk}}~\cite{lukasz2014selling-privacy-auction, imcRTB2017}. 
These companies have already invested in the \hb protocol and process early on, capitalizing on their knowledge and market share in RTB, and most publishers tend to choose these traditional big ad-partners over smaller ones.

    \begin{figure*}[t]
	\centering      	
	\begin{minipage}{0.5\textwidth}
		\centering
		\includegraphics[width=0.80\linewidth, height=0.2\textheight]{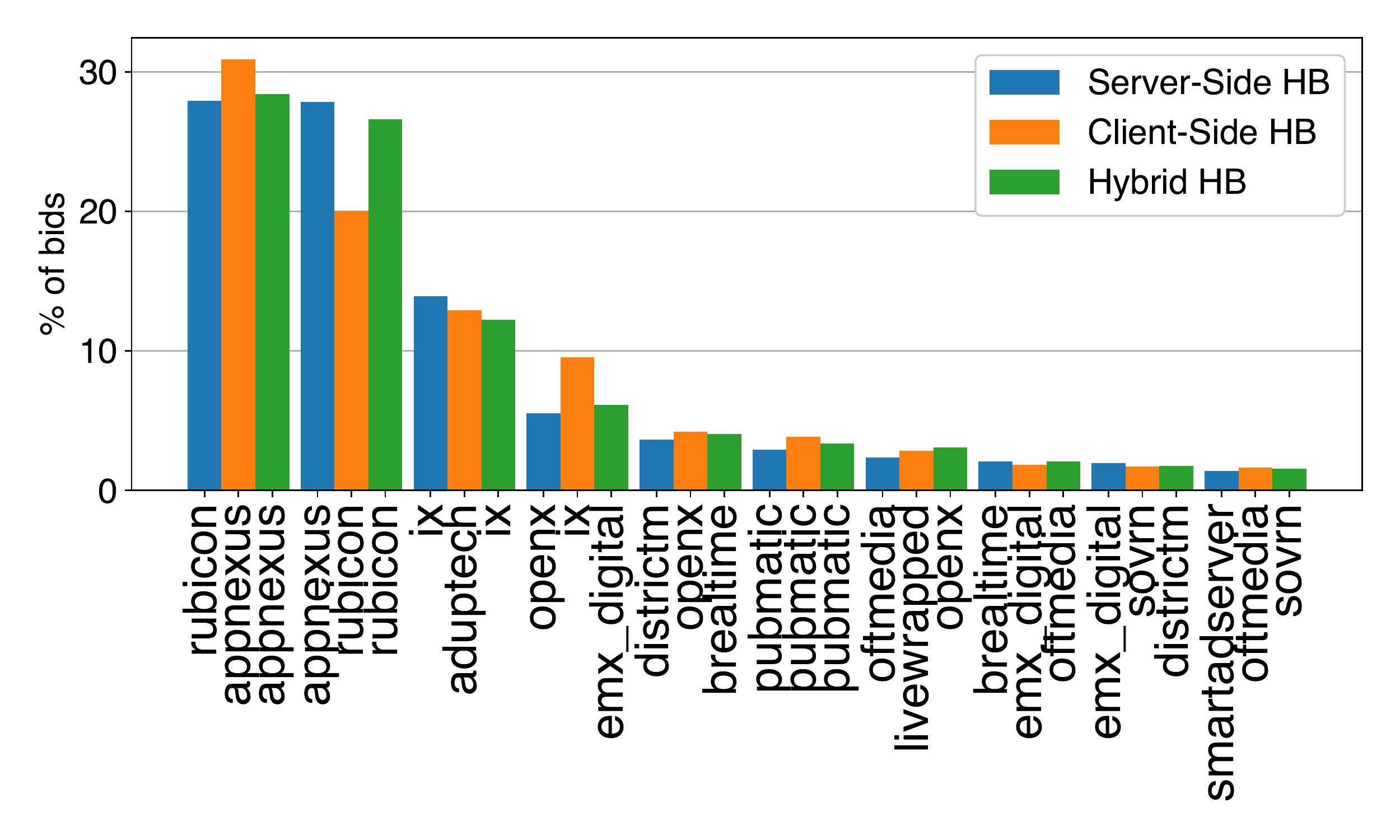}
		\caption{
			Top \DPs with respect to participation in \hb auctions, per facet of \hb.
				Top 2 partners in all types are Rubicon and AppNexus.
		}
		\label{fig:DPsPerFacet}
	\end{minipage}
	\hfill
	\begin{minipage}{0.45\textwidth}
		\centering
		\includegraphics[width=0.8\linewidth, height=0.2\textheight]{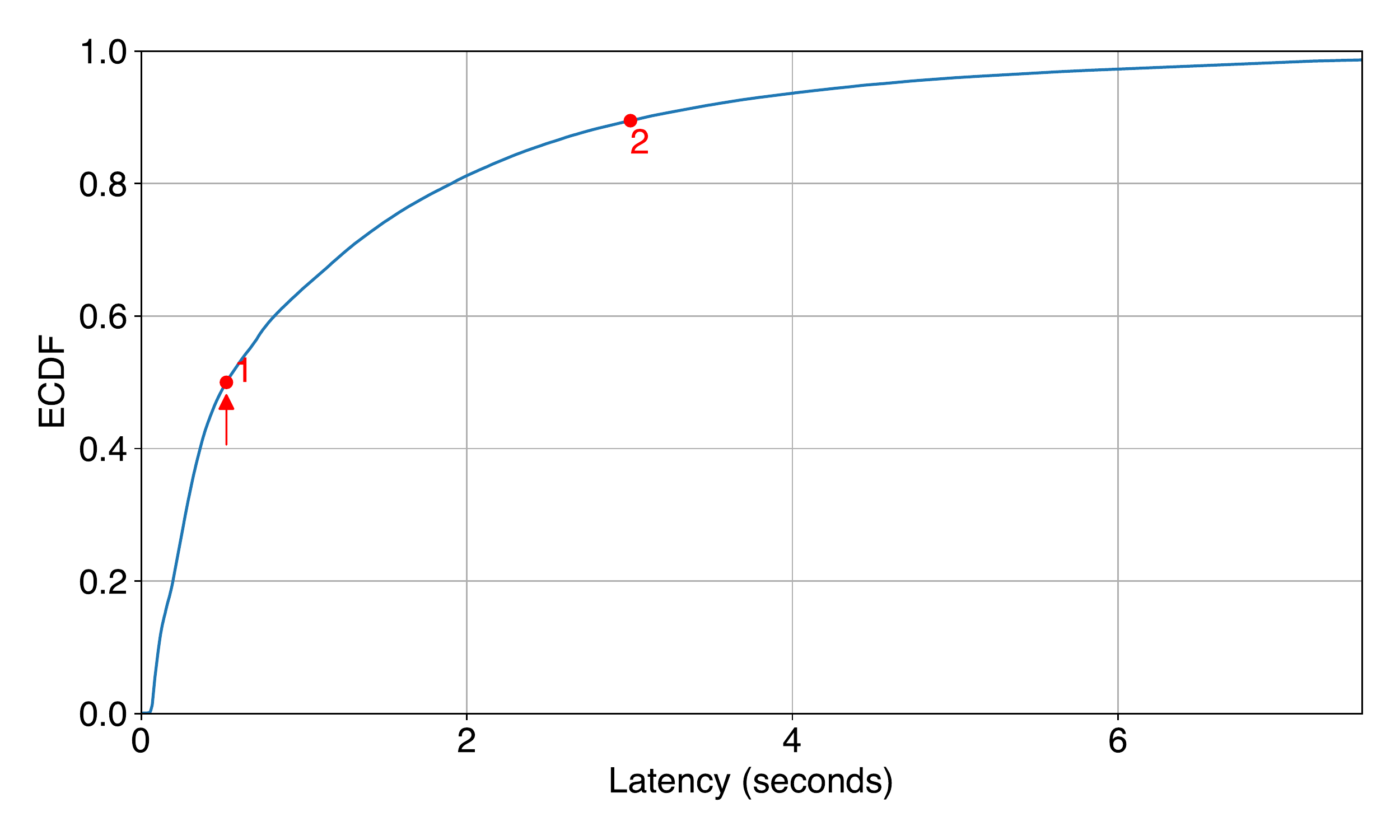}
		\caption{Total \hb latency per website. (1) Marks median latency of 600ms. (2) Marks a commonly used industry threshold of 3 seconds which captures 90\% of bid responses.}
		\label{fig:totalHB}
	\end{minipage}
\end{figure*}

	\begin{figure}
		\centering
		\includegraphics[width=0.8\linewidth, height=0.2\textheight]{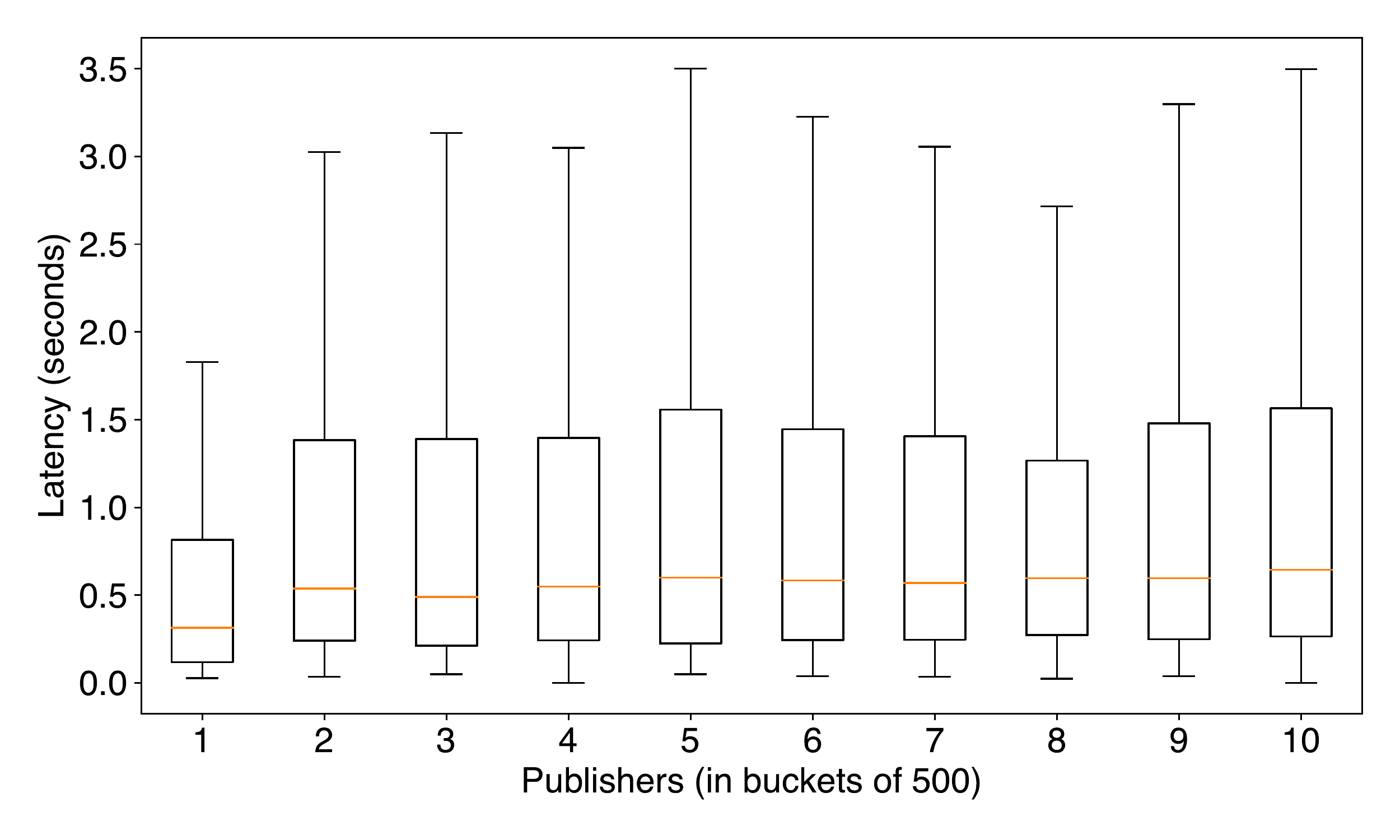}
		\caption{\hb latency Vs. domain popularity with respect to Alexa ranking (in bins of 500 websites). Some outliers that goes as high as 10 seconds (removed for clarity).} 
		\label{fig:publisher-popularity}
	\end{figure}
\noindent{\bf How many \DPs are typically used?\\}
A website can use more than one \DP during the \hb auction.
But given that the more partners used could impact the loading time of the website, a question is what is typically employed by publishers.
The number of unique \DPs participating in a \hb auction are extracted from the incoming web requests that trigger corresponding \hb events at the browser, and detected by the \toolname (see Section~\ref{sec:detection} for details on the detection).
Figure~\ref{fig:numberPrt} shows the CDF of the number of \DPs found on each website.
We can see that more than 50\% of the websites use only one \DP.
However, about 20\% of the publishers collaborate with 5 or more \DPs, and about 5\% of publishers collaborate with ten or more \DPs.

\noindent{\bf Which \DPs are typically combined?\\}
\DPs can appear on a website in different combinations.
Given that we already identified 3 types of \hb setup (client-side, server-side and hybrid), it is interesting to see how publishers select different \DPs to participate in their \hb auctions.
We should keep in mind that the mixture of partners selected can impact the performance of \hb with respect to delays and prices achieved.
Also, frequently selected combinations may reveal typical or unlike competitions between \DPs.
Therefore, we analyze the common combinations of \DPs found on webpages, and show the top 15 with respect to popularity in Figure~\ref{fig:DPcombinations}, out of 753 possible groups of competitors found.

As expected, DFP holds a majority of the market on its own (\ie appearing without any competitors) in 48\% of the cases.
Interestingly, besides DFP which dominates the market as single-partner, common groups of competitors include DFP in 51\% of groups found.
Furthermore, Criteo and Yieldlab follow as single partners with 2.37\% and 1.68\%, respectively.
Some popular pairs of competitors include DFP and other companies such as Amazon, Criteo, and AppNexus.
Finally, some triples include the above pairs with added entities such as Rubicon, OpenX, etc.

\noindent\noindent{\bf Which \DPs are used in each \hb facet?\\}
Given the three \hb facets, we anticipate that some \DPs and publishers will prefer one facet of \hb over another.
Therefore, we analyze the participation of \DPs into each type, in Figure~\ref{fig:DPsPerFacet}.
In contrast to \cshb, which all the bidders are transparent to the client, in Hybrid and \sshb this is not the case.
For this reason, we analyze the responses from the ad server (most commonly the DFP) to find the partners who won the auctions.
As expected, big DSPs like AppNexus and Rubicon hold the highest shares, followed by Index Exchange.

\begin{figure*}[t]
	\centering
    \begin{minipage}{0.38\textwidth}
		\centering
		\includegraphics[width=1.05\linewidth]{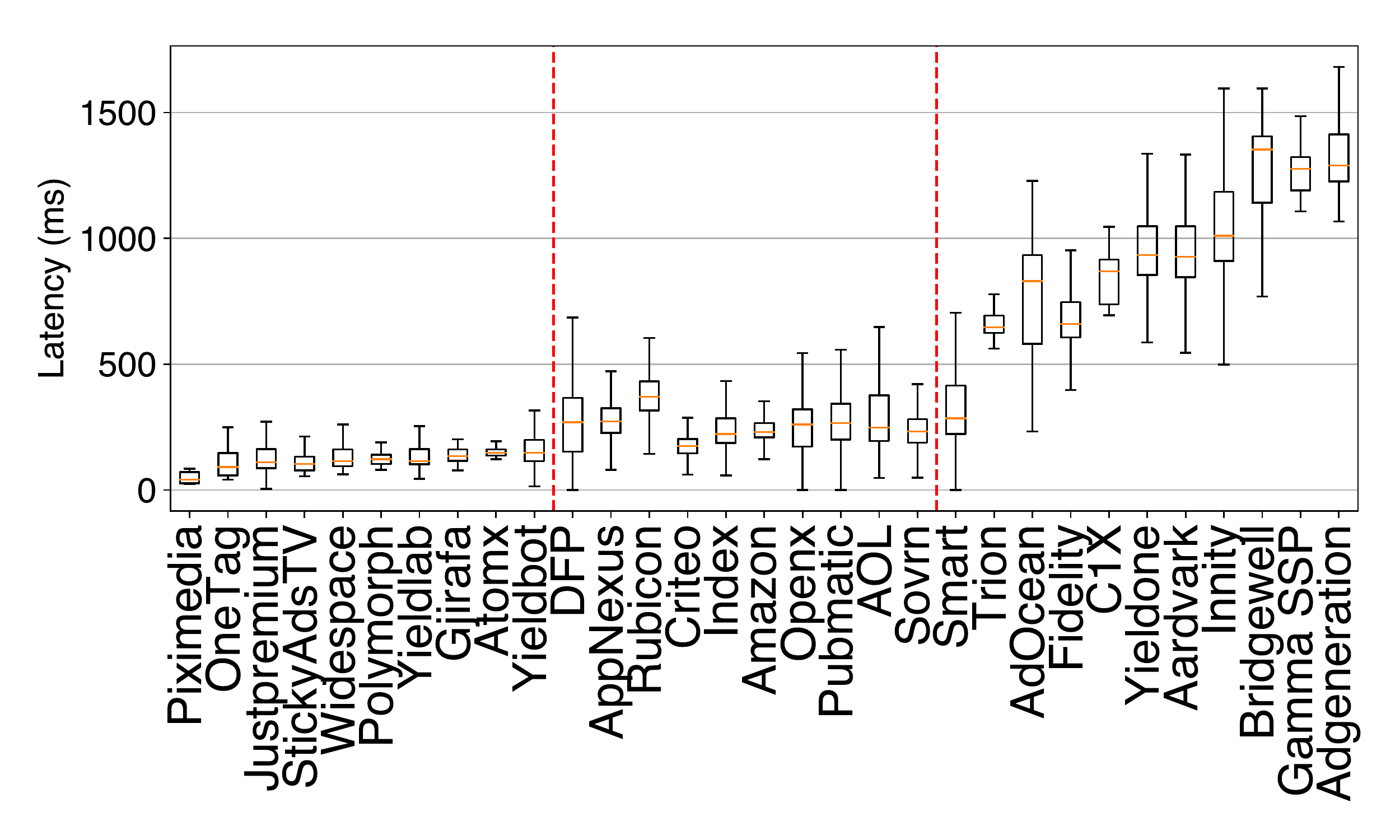}
		\caption{\hb latency for the fastest partners (left-side of plot), top partners in market share (middle-section of plot), and slowest partners (right-side of plot). Top partners in market share are not the fastest.}
        \label{fig:fast_top_slow}
    \end{minipage}
    \hfill
    \begin{minipage}{0.29\linewidth}
		\centering
		\includegraphics[width=1.05\linewidth]{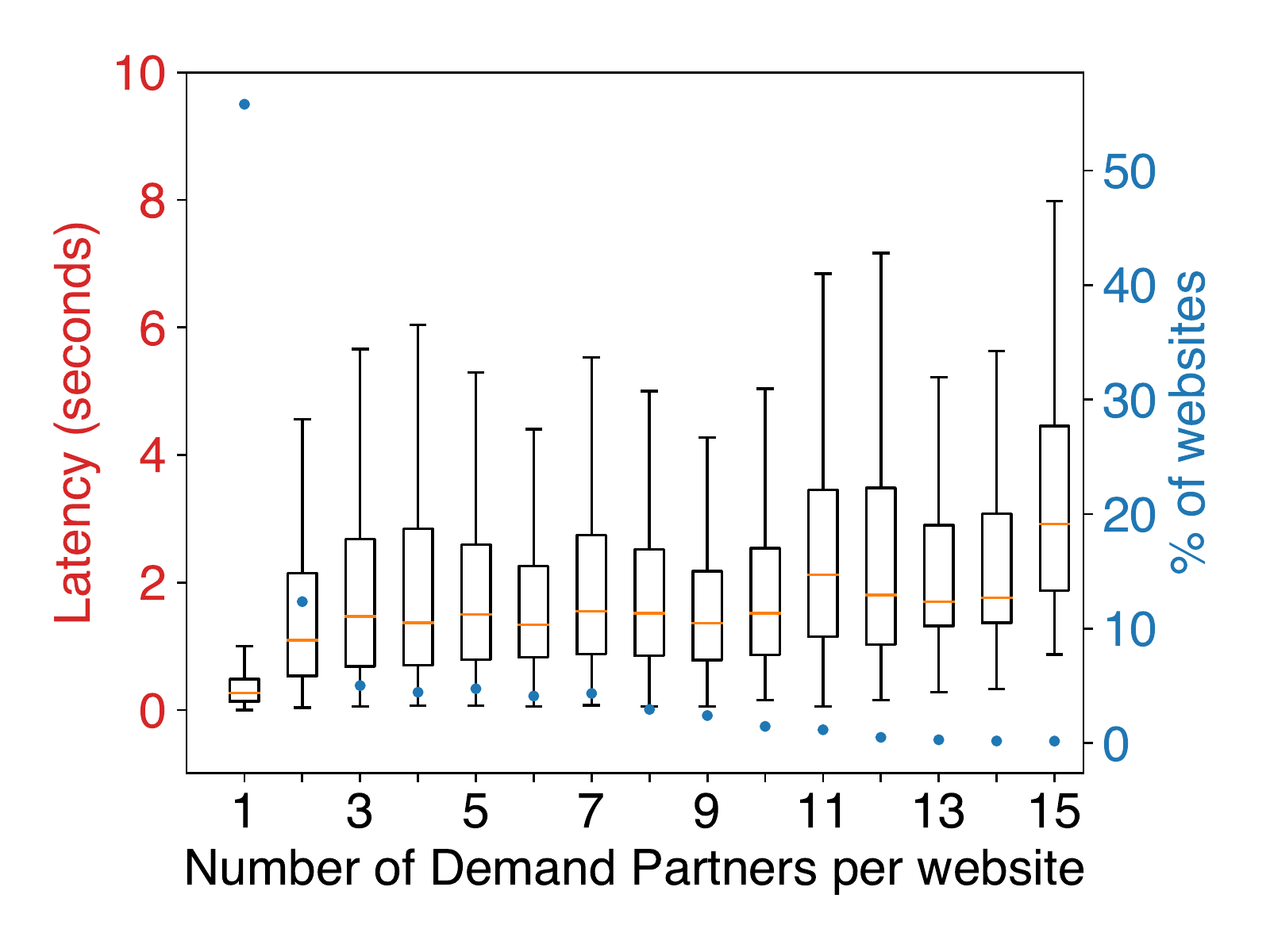}
		\caption{Total \hb latency (left y-axis) and \% of websites found (right y-axis) vs. number of \DPs per website. Publishers with more than one partner tend to have higher page load times.
		}
		\label{fig:partnerHBlat}
	\end{minipage}
    \hfill
    \begin{minipage}{0.29\textwidth}
		\centering
		\includegraphics[width=1.05\linewidth]{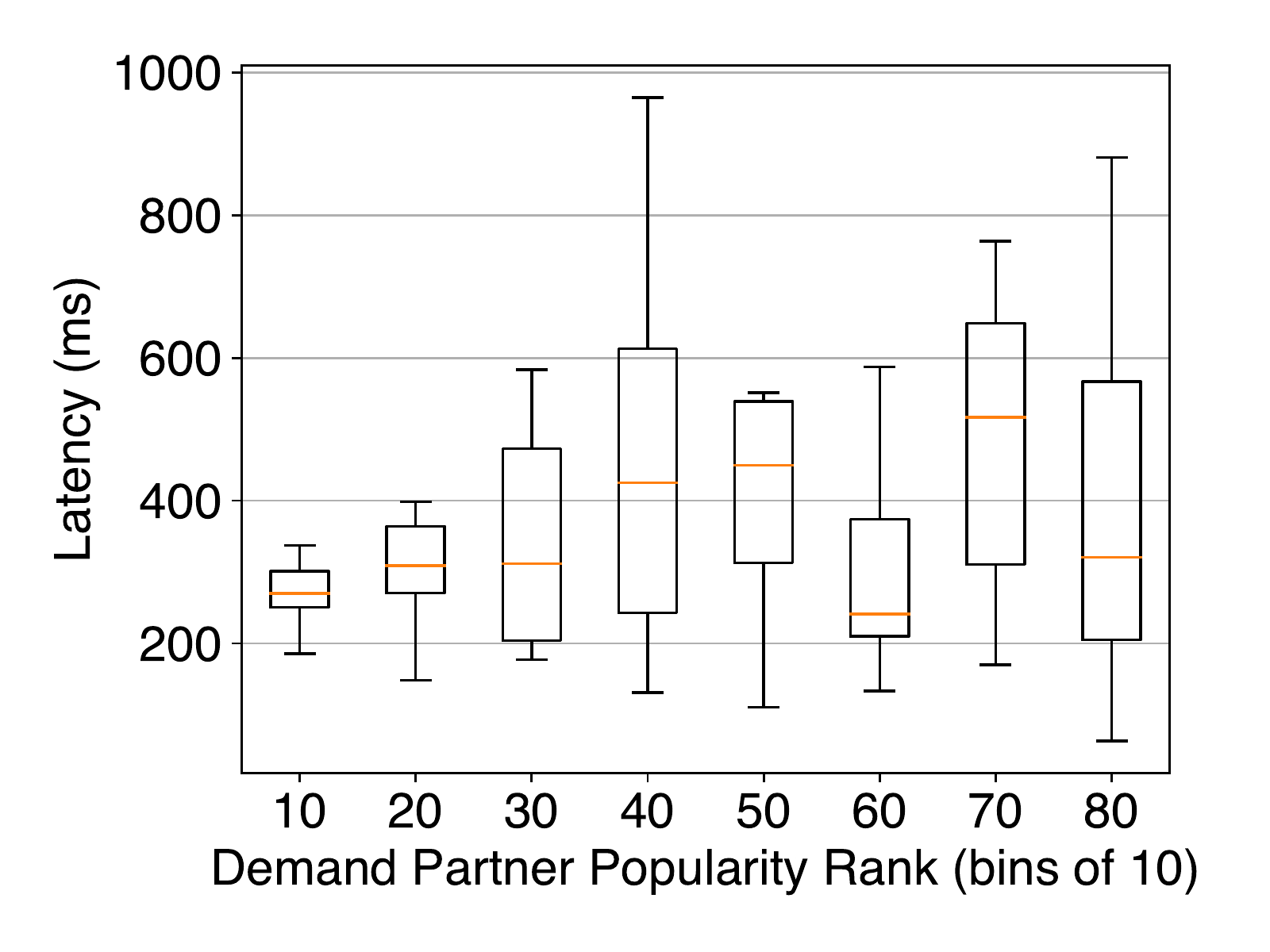}
		\caption{
		Distribution of latencies observed per \DP across all the websites.
		Partners are ranked based on popularity.
		Popular partners tend to have latencies with smaller variability.
		}
		\label{fig:prtpopularity}
	\end{minipage}
\end{figure*}

\subsection{\hblong Latency}\label{sec:latencies}

In this section, we explore various aspects of \hblong such as the imposed latency measured from different vantage points, with respect to overall latency, publishers using it, number of partners participating, etc.
In all whiskers plots, we show 5th and 95th percentiles, and the boxes show 25th and 75th percentiles, with a red line for median (50th percentile).

\noindent{\bf How much latency does \hb add?\\}
The total latency of \hb on a publisher's webpage is defined as the time from the first bid request to a \DP (step 1 in Fig.~\ref{fig:hb_UML}) until the ad server is informed and responds (step 3 in Fig.~\ref{fig:hb_UML}).
In Figure~\ref{fig:totalHB}, we show the total time needed from the \hb to process the bid requests and responses.
We see that the median latency is about 600ms (point 1 in figure).
However, some websites suffer a much higher overhead.
Indeed, about 35\% percent of the websites observe more than one second of latency, and as much as 4\% of websites observe more than 5 seconds of latency for the \hb process to conclude.

Based on our description so far, one might expect that a timeout would be used during \hb, to cut off responses from slow \DPs.
Although many of the wrappers use a timeout of 3 seconds, publishers are able to set their own threshold by making some changes in the wrappers. 
Unfortunately, our results indicate that at least 10\% of the websites exceed the threshold of 3 seconds (point 2 in Figure~\ref{fig:totalHB}), and some even need 20 seconds before the \hb is completed (not shown in the figure for clarity of the other results).

Overall, even though most of the \hb libraries strive to perform \hb activities in an asynchronous fashion, it appears that \hb can add significant overhead to a website if the library is badly tuned and \DPs are slow to respond.
In a recent report~\cite{PLT_makrM}, the average page load time (PLT) of a webpage was measured at 8.66 seconds, which is above the median latency measured here for \hb.
However, the industry recommends that the PLT should be kept under 3 seconds~\cite{PLT_makrM}, which would lead 10\% of websites with \hb auctions experiencing time delays above this recommendation.

\noindent{\bf Does publisher popularity associate with \hb latency?\\}
As a next step, we study the latency measured with respect to the ranking of each website.
Someone could expect that highly ranked publishers seek to have lower latencies for their websites, and therefore add partners in their \hb process who demonstrate lower latencies.
Also, higher-ranked websites may have available more resources to use in their \hb planning, which could lead to reduced latencies and better performance.
In Figure~\ref{fig:publisher-popularity}, we show the latency of publishers vs. their Alexa ranking.
Indeed, we find that the highest-ranked publishers (i.e., the first 500 websites) exhibit significantly lower latencies (median = 310ms), than the rest of the ranked websites (median = 500ms).

\noindent{\bf Which are the fastest and slowest \DPs?\\}

Figure~\ref{fig:fast_top_slow} shows the fastest, top and slowest \DPs, respectively.
We notice many small or unknown \DPs in these lists.
The fastest (slowest) \DPs have median values in the range of 41-217ms (646-1290ms).
Interestingly, the top \DPs with respect to market share have latencies that are small, but not low enough to qualify them for the fastest partners (with the exception of Criteo which has a median latency under 200ms).

\noindent{\bf Do multiple \DPs impact \hb protocol's latency?\\}
As we mentioned earlier, a publisher may choose to use several \DPs at the same time.
Although this decision may increase competition for the ad-slots offered, and can drive-up the bidding prices, and consequently the publisher's revenue, it may also increase the latency of the webpage to load on the user's browser, and decrease the quality of the overall user experience. 
Therefore, we explore the impact that the number of \DPs can have on the user experience with respect to latency.

Figure~\ref{fig:partnerHBlat} shows the latency of websites vs. the number of \DPs each website has.
We observe that publishers who use only one \DP have a small latency of 268.2 ms.
As can be seen by the second y-axis, this is the majority of websites.
Also, publishers with two \DPs have a  latency of 1091.6 ms.
Publishers with more than two \DPs have a median latency in the range of 1.3-3.0 seconds.
\noindent{\bf Does \hb partner popularity associate with \hb latency?\\}
Next, we study the latency of all 84 \DPs detected, ranked based on their popularity in our dataset.
In Figure~\ref{fig:prtpopularity}, we show the distribution of latencies observed per partner, when computed across all the websites each partner was found.
We observe that the most popular partners tend to have latencies with smaller variability (up to 200ms), in comparison to the less popular partners who may exhibit latency variability up to 500-1,000ms.

\begin{figure*}[t]
    \centering
    \begin{minipage}{0.3\textwidth}
		\centering
		\includegraphics[width=1.1\linewidth, height=0.18\textheight]{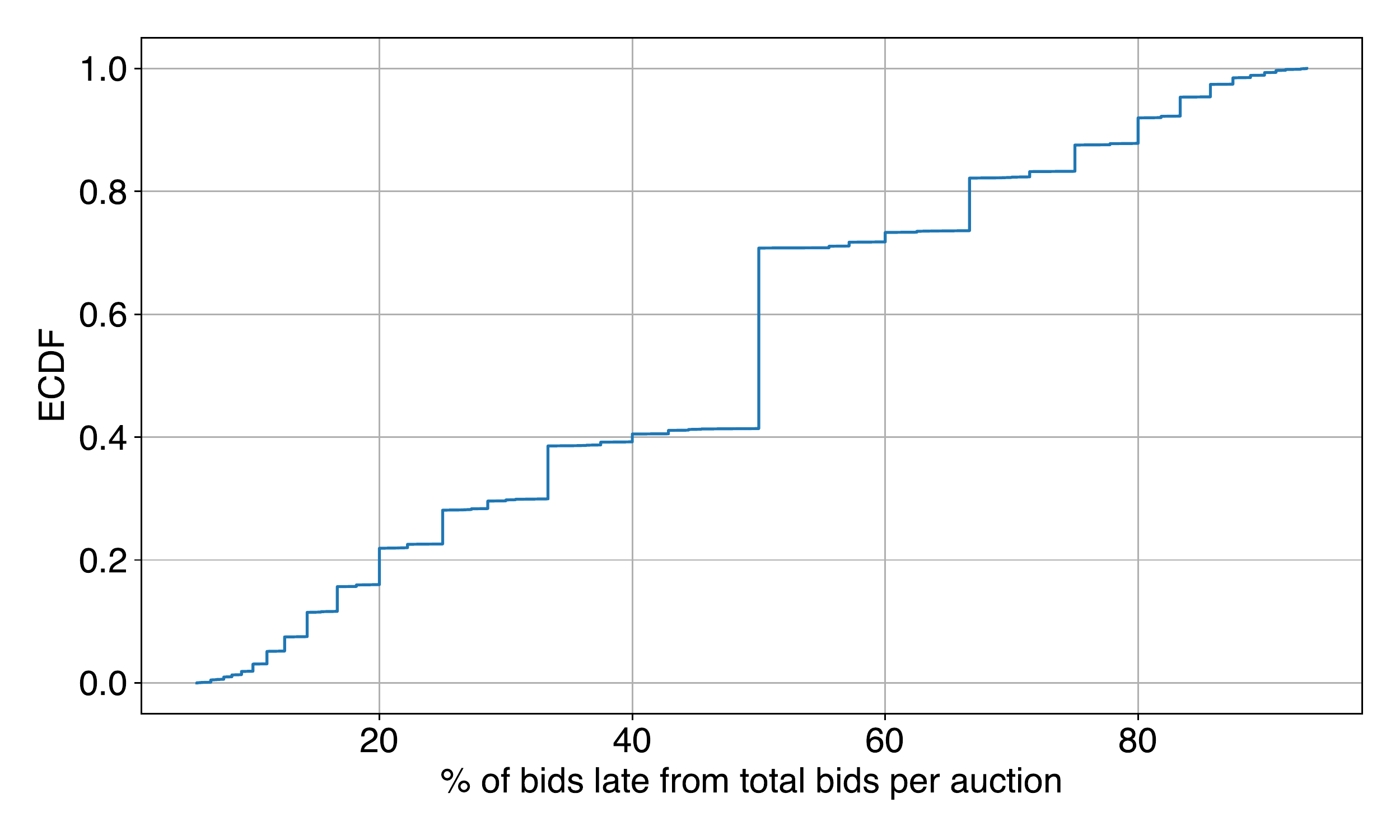}
		\caption{Portion of late bids over the total bids received, due to high latency of a partner to respond.
		For 10\% of auctions, they have 80\% or more late bids.}
		\label{fig:lostbids}
	\end{minipage}
    \hfill
    \begin{minipage}{0.35\textwidth}
		\centering
		\includegraphics[width=1.1\linewidth]{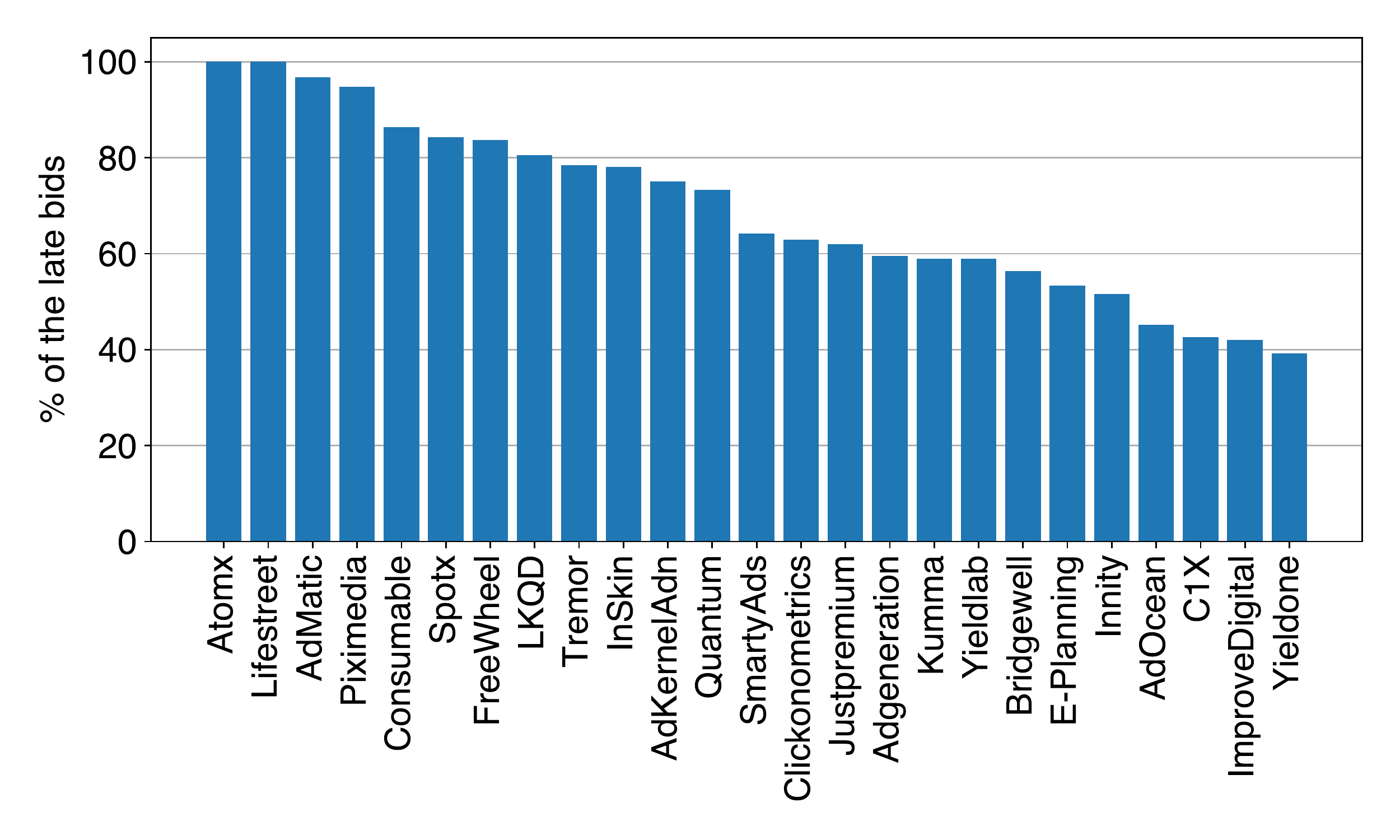}
		\caption{
		Percentage of late bids out of all bids sent per \DP.
		Some partners have all their bids arriving too late to be considered for auction.}		
		\label{fig:LateBidsPerDmprt}
	\end{minipage}
    \hfill
    \begin{minipage}{0.3\textwidth}
        \centering
	    \includegraphics[width=1.1\linewidth]{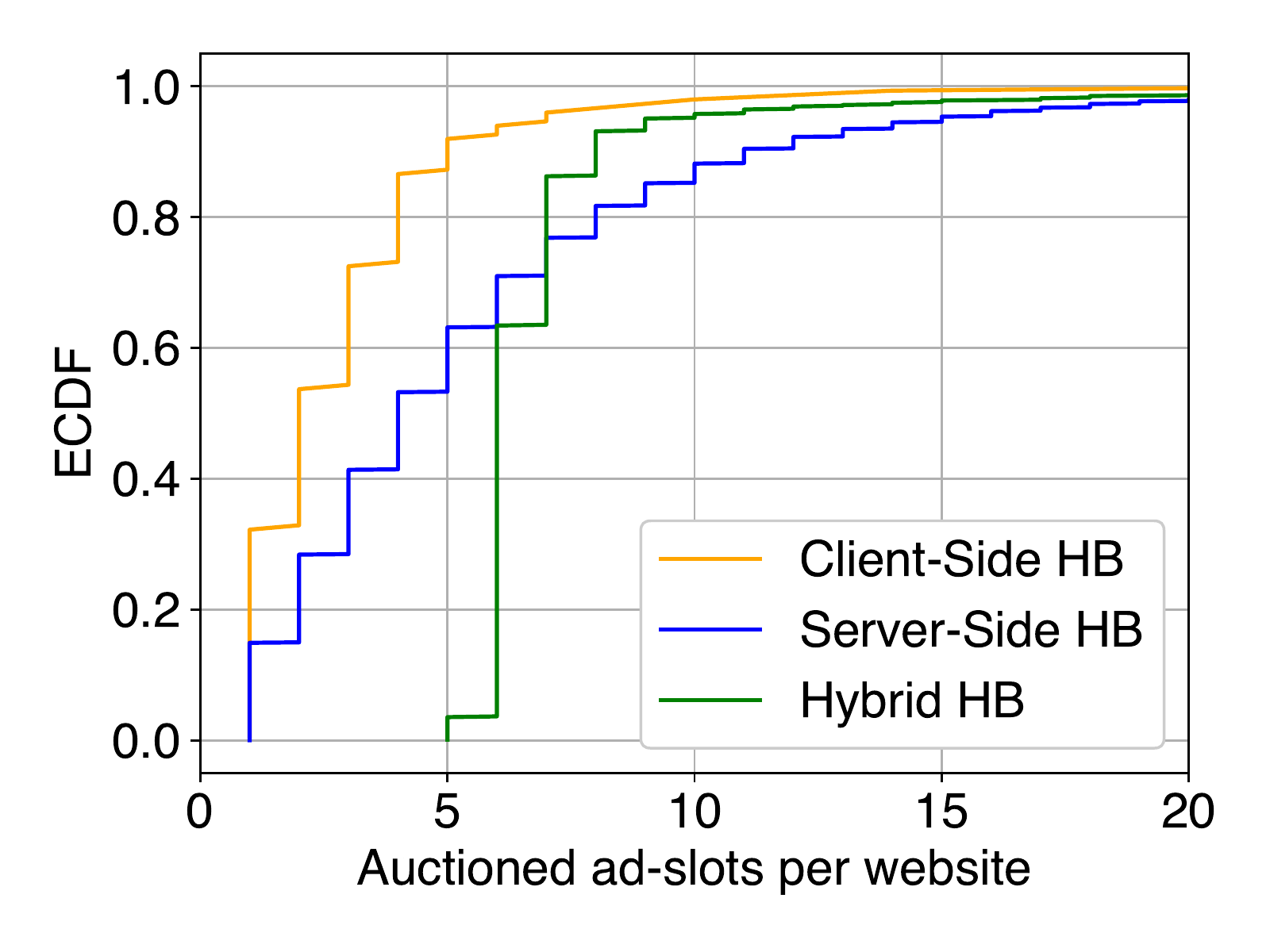}
	    \caption{
    	Auctioned ad-slots across websites, per \hb facet.
	    The median website has 2-6 available ad-slots. 90\% of websites have 5-11 ad-slots (depending on the \hb type).
	    }      
	\label{fig:auctionedSlots}
    \end{minipage}
\end{figure*}

\noindent{\bf How many bids are late?\\}
Here, we analyze the portion of bids that arrive too late to be included in the auction.
As late bids we define all the responses about bids from \DPs which arrive too late, \ie \emph{after} the request to the ad server is sent from the browser.
Thus, it is important to understand what is the portion (and number) of bids that were received from the browser, that came too late and were not considered in the \hb auction.
In Figure~\ref{fig:lostbids}, we show the CDF of the portion of such late bids with respect to the total number of bids received at a website for the \hb auction.
We see that in 50\% of the cases with late bids, almost 50\% of the bid responses come too late to be considered in the auction by the ad server.
Also, for 10\% of the auctions, more than 80\% of the bids are late.
In results not show here due to space,
we measured that in 60\% of the auctions, there was only one late bid, in 40\% of the auctions there was at least two late bids, and in 20\% of auctions at least four late bids.

In Figure~\ref{fig:LateBidsPerDmprt}, we measure the percentage of late bids per \DP.
We notice that 21 \DPs bid too late in 50\% of the auctions they participate. 
In some extreme cases, the \DP loses 100\% of the bids they send.
All these late bids point to the possible loss of revenue from the publisher.
This could be the result of a poorly defined wrapper that sends the request to the ad server the same time it sends the requests to \DPs, without waiting for their responses first, as well as \DPs that do not have the proper infrastructure to respond fast enough to all incoming requests.

\begin{figure*}[t]
	\centering
	\begin{minipage}{0.45\textwidth}
		\centering
		\includegraphics[width=1.0\linewidth, height=0.25\textheight]{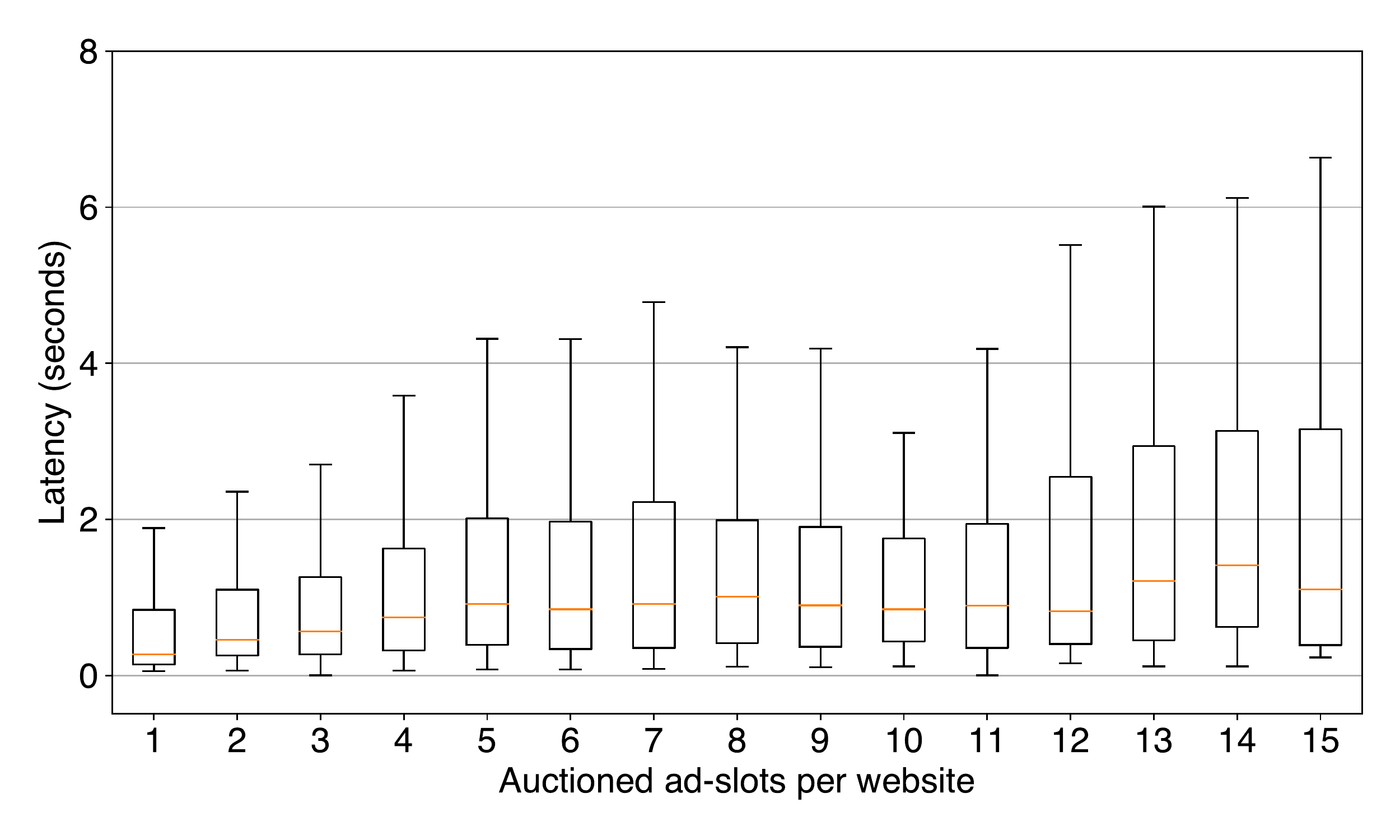}
		\caption{\hb latency as a function of the number of ad-slots auctioned.
			More ad-slots result in higher median latency and variability in latency for the \hb process.}
		\label{fig:adLatency}
	\end{minipage}
	\hfill
	\begin{minipage}{0.48\textwidth}
		\centering
		\includegraphics[width=1.0\linewidth]{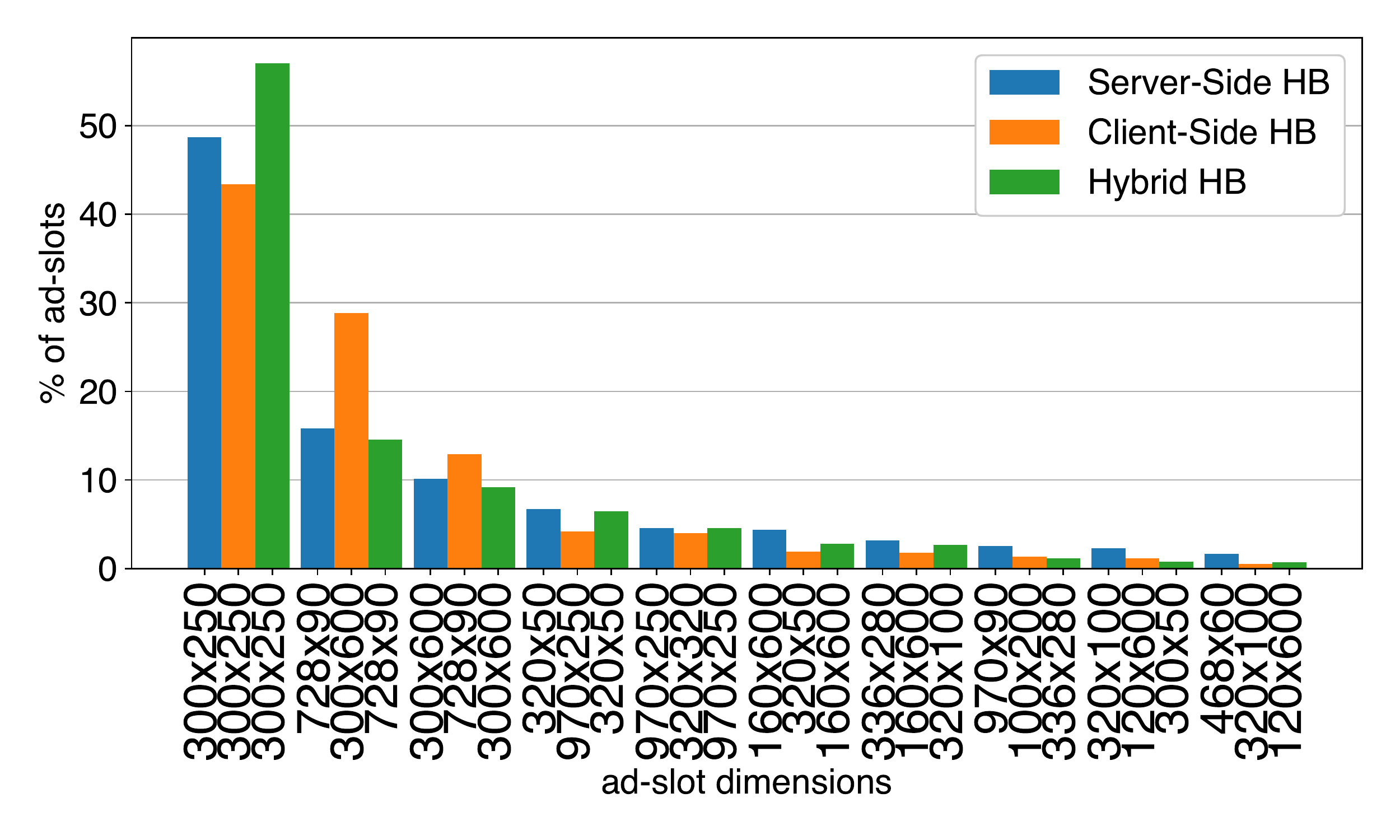}
		\caption{
		Portion of ads for different \hb ad sizes, per \hb facet.
		The side banner (300x250) and top banner (728x90) are among the most popular ad-slot sizes in all \hb facets.
		}
		\label{fig:adDim}	
	\end{minipage}

\end{figure*}

\subsection{Auctioned Ad-slots}\label{sec:ad-auctions}
In this section, we investigate the properties of the auctioned ad-slots, such as the size, the number of auctions per website, and how this impacts the overall performance of the protocol.

\noindent{\bf How many ad-slots are auctioned per webpage?\\}
We start by investigating the number of ad slots that are available for auction.
In Figure~\ref{fig:auctionedSlots}, we plot the CDF of the number of ad-slots across the websites, per \hb type.
In general, and for up to 70\% of websites, the Hybrid \hb type auctions more ad-slots than the other two types.
For the other 30\% of websites, \sshb auctions more ad-slots.
The median website has 2-6 available ad-slots, and 90\% of websites have up to 5-11 ad-slots (depending on the \hb type).
Also, 3\% of the websites provide more than 20 slots for auction.

Requesting bids for 20 ad-slots on a single page can be considered odd, even a flag for suspicious or fraudulent behavior.
Therefore, we manually investigated such cases, and to our surprise, we found that some publishers request auctions for \emph{more} slots than they have available for display.
After investigating this behavior further, we observed that these ad-slots refer to several different devices and screen sizes such as for tablet, smartphone, laptop, etc.
We speculate they do that due to either bad configuration of their wrapper (\ie they use the same \hb wrapper for all the devices they serve without customizing the requests), or because they want to receive bids for multiple versions of the same ad-slots, for better optimization of the publisher's \hb process later on.
Indeed, this odd activity needs to be studied in depth in the future, to understand if it is a matter of bad practice or an effort for ad-fraud.

\noindent{\bf Does the number of auctioned ad-slots impact latency?\\}
Next, we checked if the \hb latency is associated with the number of ad-slots auctioned.
Intuitively, we may expect that the more slots are to be auctioned, the more time the \hb will take.
However, given that a lot of \DPs invest significant computing resources to parallelize and optimize bidding computations, the above statement may not hold.
In Figure~\ref{fig:adLatency}, we plot the latency of \hb based on the number ad-slots auctioned in the website.
In the majority of cases, this latency includes the communication to the ad server.
In \cshb, we cannot know the ad server (since each publisher uses their own), so we have no means to infer this latency.
We observe that the total latency tends to increase with the number of slots auctioned.
In fact, when there are 1-3 ad-slots auctioned, the median latency is 0.30-0.57 seconds, but when the slots are 3-5, the median latency ranges to 0.57-0.92 seconds.
Interestingly, we observe that even if there is only one ad-slot to be auctioned, the latency can still vary per auction, from a few tens of milliseconds to almost two seconds.
This variability can be due to extra latencies as result of internal auctions occurring at each \DP.

\noindent{\bf What are the most popular ad-slots auctioned?\\}
Finally, we analyze the most popular dimensions of \hb ad slots. 
Our findings are presented in Figure~\ref{fig:adDim}, per facet of \hb.
The most common ad size is the 300x250 (side banner), for all 3 facets.
The second most common is the 728x90 (top banner) and the 300x600 (for the \cshb).
These are generally popular banners in both mobile and desktop advertising, and they match results observed in the past for RTB~\cite{imcRTB2017}.
Due to the increase of mobile browsing, publishers can choose these specific sizes to keep the \hb configuration simple and well defined for multiple devices (as they don't need to set multiple sizes for different devices, and fewer auctions need to occur on the \DPs' side).

\subsection{Ad-slot Bid Prices}\label{sec:ad-prices}
In this section, we discuss the auctioned ad-slots' bid prices and how they vary depending on their size. 
We were able to detect the ad prices using \toolname.
In case of Hybrid and \cshb, most of the prices are transparent to the client and easy to extract from the \textit{bid response} messages.
In contrast, in \sshb the prices are not trivial to detect.
We analyze in depth the auction metadata, and based on several heuristics we find and extract the prices whenever they are included.

\begin{figure}
	\centering
	\includegraphics[width=0.8\linewidth, height=0.2\textheight]{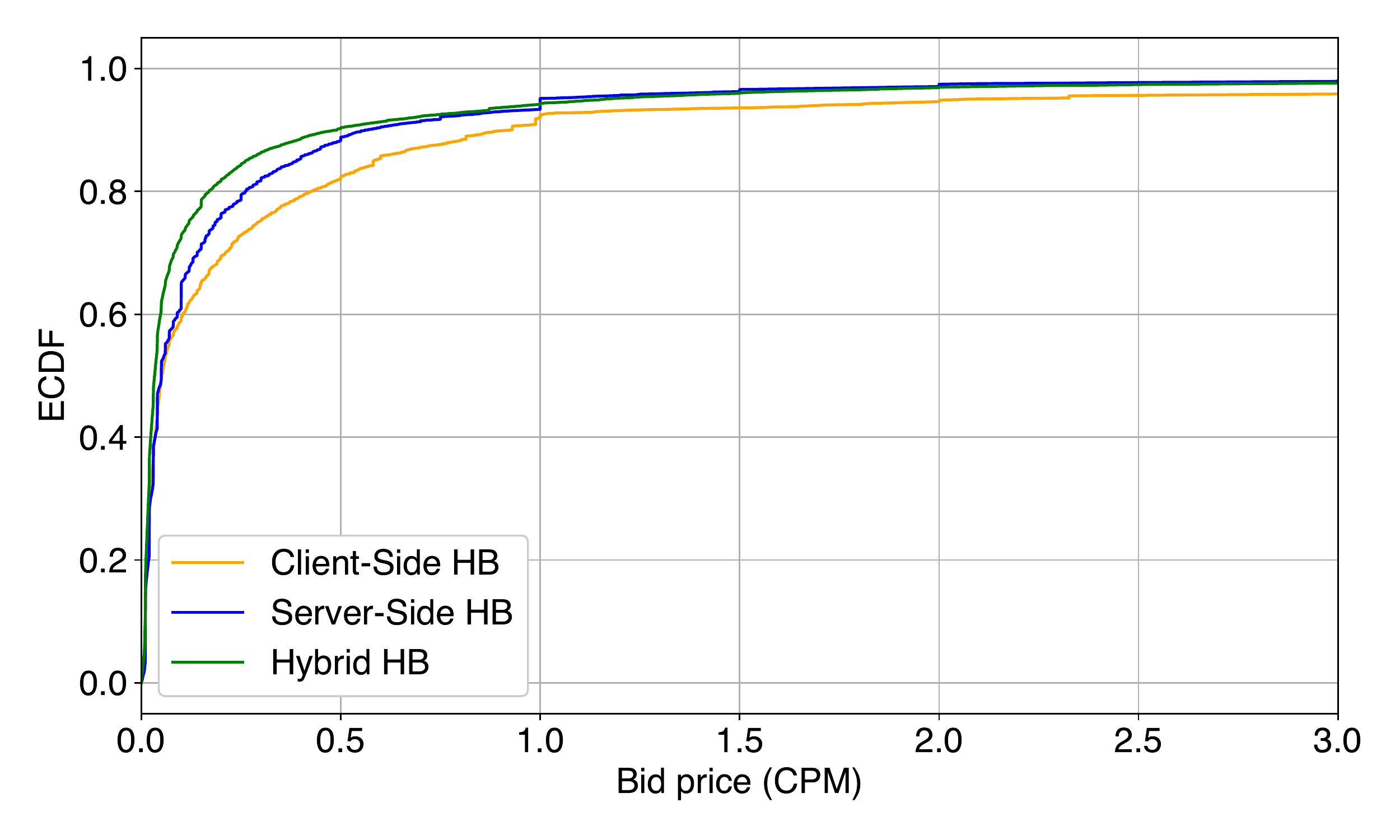}
	\caption{CDF of the auctioned ad slots bid prices, per \hb facet.
	These are baseline prices that \DPs are willing to spend when they have no information for the user.}
	\label{fig:ecdfPrices}
\end{figure}

\noindent{\bf What are the \hb partners willing to pay?\\}
First, we analyze the prices bided by the \DPs during the auctions.
In Figure~\ref{fig:ecdfPrices}, we show the CDF of the baseline bid prices (in CPM or cost per thousand ad impressions, in USD) that advertisers are willing to spend for the ad-slots auctioned, per type of \hb.
In general, we note that \cshb draws higher bid prices for the publisher, in comparison to the other two types.
Also, more than 20\% of the prices are more than 0.5 CPM, which is lower but comparable to regular waterfall prices, as claimed in past studies (found to be $\sim$1 CPM~\cite{imcRTB2017}).
Also, we should note that these prices are baseline, so they are much lower than if they were referring to targeted users.

\begin{figure}[t]
	\centering
	\includegraphics[width=.8\linewidth, height=0.2\textheight]{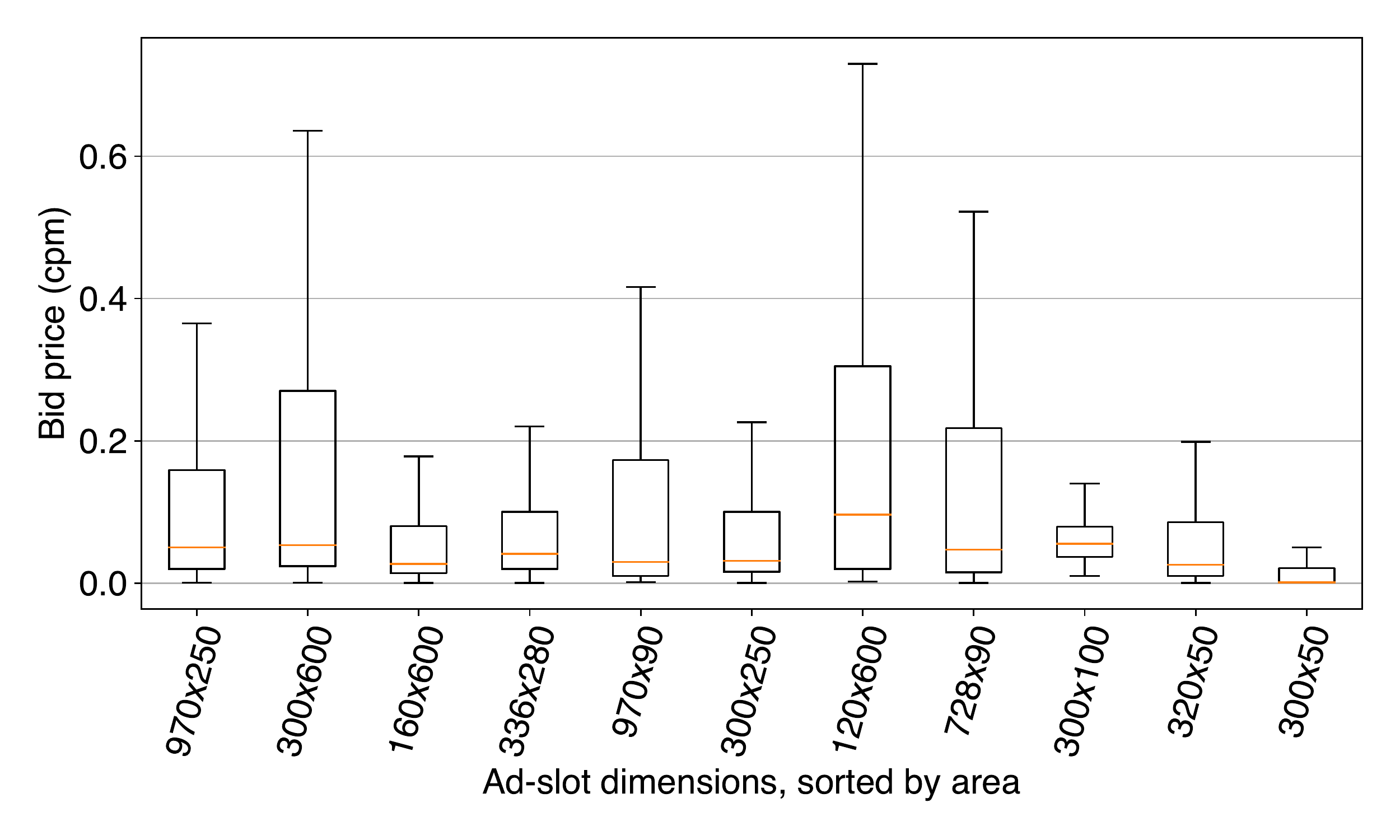}\vspace{0cm}
	\caption{
		Distribution of bid prices in CPM per ad-slot size (x-axis sorted by area of ad-slot).
		Even in our crawler's baseline scenario, partners bid high prices to reach users.}
		\label{fig:cpmsize}
\end{figure}

\noindent{\bf What are the \hb partners paying per ad-slot?\\}
Second, we compare ad-slot sizes with bid prices for each size.
In Figure~\ref{fig:cpmsize} we plot the prices (in CPM) for each ad-slot.
We see that in the recorded dimensions, the median cost ranges from 0.00084-0.096 CPM.
The most expensive ad-slot (based on median price) is 120x600 with 0.096 CPM.
The cheapest ad-slot is 300x50 (which also happens to have the least ad-area) with 0.00084 CPM.
Also, the most popular ad-slot size, which is 300x250, has a median cost of 0.031 CPM.
Previous studies on waterfall standard~\cite{imcRTB2017} find the prices of 300x250 slot ranging from 0.1 to 1.4 CPM, with a median of 0.19 CPM.
These prices are higher than the ones found in our \hb study, but we should again consider that our detected prices are for baseline users that \DPs have no prior knowledge, whereas in~\cite{imcRTB2017} it was for real users.
Therefore, a follow-up work could apply real user profiles to collect \hb prices, and thus, make a more fair comparison with RTB prices.

\begin{figure}
    \centering
    \includegraphics[width=0.8\linewidth, height=0.2\textheight]{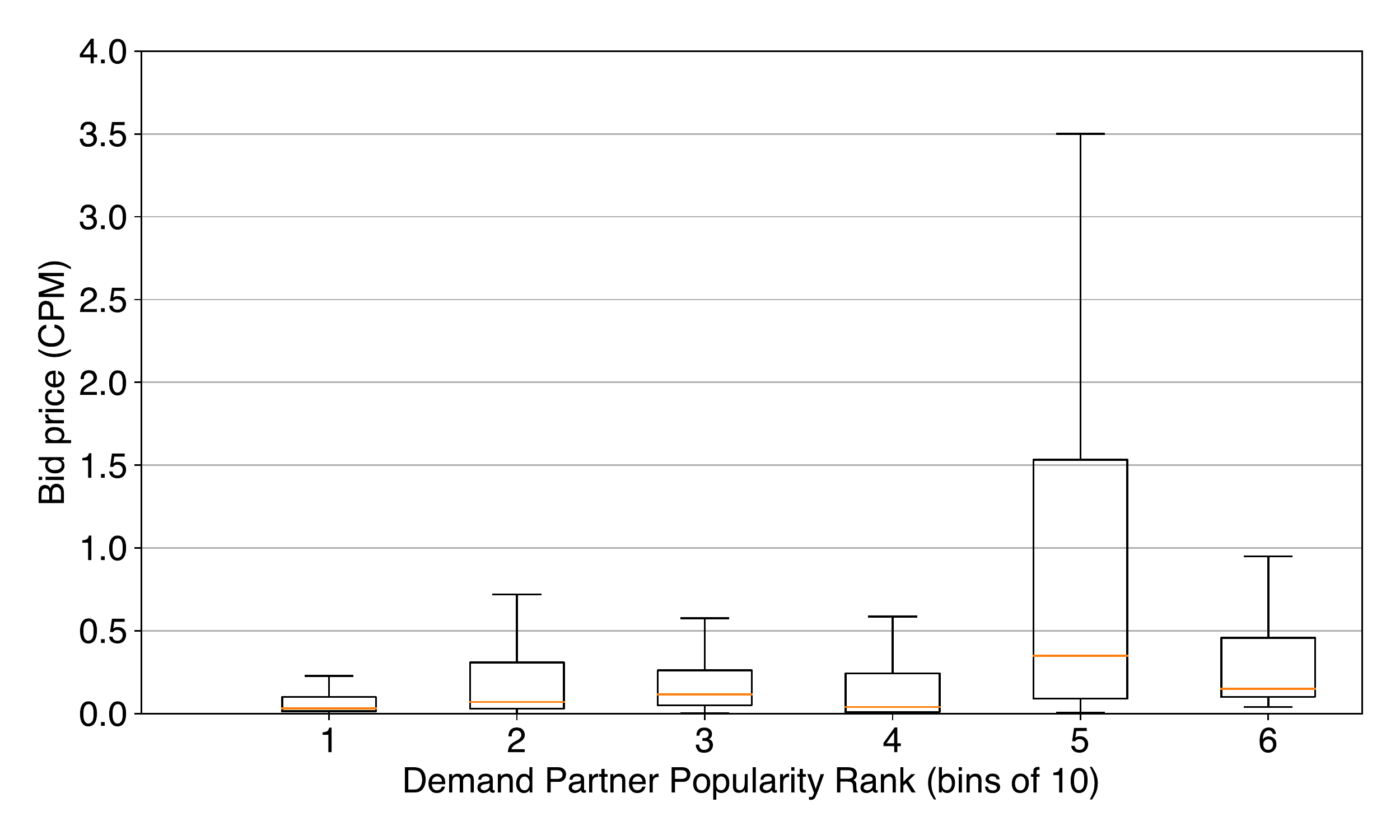}
    \caption{Distribution of prices that partners bid, ranked by popularity of \DP (who are grouped in bins of 10).}
    \label{fig:variability}
\end{figure}

\noindent{\bf What is the variability of bid prices per DSP?\\}
In Figure~\ref{fig:variability}, we plot the prices (in CPM) that each \DP bid to examine possible association between a partner's popularity and how high they bid in \hb.
The DSPs are ranked by popularity and grouped in buckets of 10 to ease illustration.
The most popular partners (first bins) tend to be more consistent and bid lower prices.
In contrast, less popular DSPs have higher median bid prices and variability in their bids.
This observation could be explained when considering how the \hb market works: for less popular DSPs to be competitive and win auctions, they bid higher prices than popular partners to reach sufficient number of users.
Alternatively, this result could also indicate that more popular partners have technology that detects when browsing is of a baseline (or bot/unknown) user and therefore do not bid high, whereas the less popular partners bid high, hoping to target a real user.
Finally, it can also be a side effect of how \DPs decide to spend their budget across the websites they collaborate with: more popular partners exist in more websites, and may chose to bid low in many of them, to cover a wider range of websites.

\section{Related Work}

User data and their economics have long been an interesting topic and attracted a considerable body of research~\cite{pachilakis2019measuring,imcRTB2017,fdvt,bigmac,staiano2014moneywalks,acquisti2013privacy,forSale,lukasz2014selling-privacy-auction,vallina2016tracking,followTheMoney,Zarras:2014:DAM:2663716.2663719,razaghpanah2018apps}.
In particular, Acquisti \etal discuss the value of privacy after defining two concepts (i)~\emph{Willingness To Pay}: the monetary amount users 
are willing to pay  to  protect  their  privacy, and (ii)~\emph{Willingness  To  Accept}: the compensation that users are willing to accept for their privacy loss~\cite{acquisti2013privacy}.
In two user-studies~\cite{bigmac,staiano2014moneywalks} authors measure how much users value their own offline and online personal data, and consequently how much they would sell them to advertisers.
In~\cite{forSale},  authors propose ``transactional'' privacy to allow users to decide what personal information can be released and receive compensation from selling them.

Papadopoulos \etal set out to explore the cost advertisers pay to deliver an ad to the user in the waterfall standard and RTB auctions~\cite{imcRTB2017}.
In addition, they study how the personal data that users leak while browsing (like location and interests) can affect the pricing dynamics.
The authors propose a methodology to compute the total cost paid for the user even when advertisers hide the charged prices. Finally, they evaluate their methodology by using data from a large number of volunteering users.
Olejnik \etal perform an analysis of cookie matching in association with the RTB advertising~\cite{lukasz2014selling-privacy-auction} .
They leverage the RTB notification URL to observe the charge prices and they conduct a basic study to  provide some insights into these prices, by analyzing different user profiles and visiting contexts.
Their results confirm that when the users' browsing histories are leaked, the charge prices tend to be increased.
In~\cite{papadopoulos2018cost}, the authors measure the costs of digital advertising on both the user's and the advertiser's side in an attempt to compare how fairly these costs are distributed between the two.
 In particular, they compare the cost advertisers pay in the waterfall standard with the costs imposed on the data plan, the battery efficiency and (by using cookie synchronization~\cite{papadopoulos2019cookie,Acar:2014:WNF:2660267.2660347,Papadopoulos:2018:ECM:3193111.3193117} as a metric) the privacy of the specific user.

In~\cite{optimiz}, the authors briefly describe \hb\ and focus on optimizing its bidding strategy and
the produced yield. They consider revenue optimization as a contextual bandit problem, where the context consists 
of the information available about the ad opportunity, such as properties of the internet user or of the provided ad slot.
In~\cite{followTheMoney}, authors use a dataset of users' HTTP traces and provide rough estimates of the relative value of users by leveraging the suggested bid amounts for the visited websites, based on categories provided by the Google AdWords.
FDTV~\cite{fdvt} is a plugin to inform users in real-time about the economic value of the personal information associated to their Facebook activity.
In \cite{iwnder}, Iordanou \etal try to detect both programmatic and static advertisements in a webpage, using (i) a crowdsourcing, and (ii) a crawling approach to determine the criteria with which ads are displayed.
They find biases on ads depending on age, income and gender of users.

Bashir \etal study the diffusion of user tracking caused by RTB-based programmatic ad-auctions~\cite{DiffusionofUserTrackingDataintheOnlineAdvertisingEcosystem}.
Results of their study show that under specific assumptions, no less than~52 tracking companies can observe at least~91\% of an average user's browsing history. 
In an attempt to shed light upon Facebook's ad ecosystem, Andreou \etal investigate the level of transparency provided by the mechanisms ``Why am I seeing this?'' and Ad Preferences Page~\cite{andreou2018investigating}.
The authors built a browser extension to collect Facebook ads and information extracted from these two mechanisms before performing their own ad campaigns and target users that used their browser extension.
They  show that ad explanations are often incomplete and misleading.
In~\cite{bashirtracing}, the authors aim to enhance the transparency in ad ecosystem with regards to information sharing, by developing a content agnostic methodology to detect client- and server- side flows of information between ad exchanges and leveraging retargeted ads.
By using crawled data, the authors collected 35.4K ad impressions and identified 4 different kinds of information sharing behavior between ad exchanges. 

\section{Summary \& Discussion}

\hblong is gaining popularity among Web publishers, who want to regain the control of their ad inventory and what advertisers are paying for it.
Proponents of \hb have touted that this new ad-tech protocol increases transparency and fairness among advertisers, since more partners can directly compete for an ad-slot.
\hb, in theory, can boost the revenue of publishers, who can select the \DPs that are competing for the publishers' ad-slots, and also remove intermediaries from the ad-selling process.

In this study, we investigate and present in full detail the different implementations of \hb and how each of them works.
Based on these observations, we design and implement \toolname: a first of its kind tool to measure in a systematic fashion the evolving ecosystem of \hb, its performance and its properties.
By running  \toolname across a list of top \alexaTop Alexa websites, we collected data about \datasetAuctionSize \hb auctions and performed the first in-depth analysis of \hb.
We discuss our lessons from this study in the next paragraphs.

\subsection{Commoditization of Ad Supply}

\hblong was introduced to put \DPs under pressure for more competitive pricing (and loosen Google's grip on the market).
Indeed, it has changed the hierarchy on the supply side.
Depending on the publisher's needs, we found that \hblong can be implemented in 3 ways: (i) \cshb, (ii) \sshb, and (iii) Hybrid \hb.
Therefore, \DPs that could previously claim exclusive access to a publisher's inventory (and thus higher positions in the waterfall) are no longer able to do so.
Instead, \hb enabled all \DPs regardless of their size or relationship with publishers, to compete for the same inventory, thus commoditizing supply~\cite{commoditization}. 

However, as measured in this study, big companies such as DoubleClick, AppNexus, Rubicon, Criteo, \etc, took advantage of their existing dominance in the ad-market and placed themselves again in a very centralizing (and process controlling) position within the \hb ecosystem (especially within the \sshb and Hybrid \hb models).
In fact, we identified that \sshb dominates this market with 48\% of auctions handled by a single partner/ad server.
Google, in particular, handles as much as 80\% of \hb auctions.
DoubleClick for Publishers (DFP) dominates as a single partner, while it also appears in 51\% of the competing groups of \DPs in \hb.
Also, most publishers use only one \DP, but some use many (more than 10).
Interestingly, this centralization is in direct contrast to the publishers' revenue.
We found that websites utilizing \cshb achieve higher bid prices than the other two models.

\subsection{Non-Viable Performance Overheads}

The fear of latency has kept some premium publishers away from header integrations and continues to make others wary about embracing \hb.
Results of this study verify the concerns of publishers~\cite{hb-latency,hb-latency2} regarding the latencies imposed on the user side.
We measured up to 0.6 seconds for the median website and more than 3 seconds in 10\% of websites.
Furthermore, publishers with more than one \DP experienced higher \hb latencies: one \DP imposes a small latency of 0.3 seconds, but 2 \DPs impose 1.1 seconds latency, and 3 \DPs can impose up to 3 seconds latency.
It is of no doubt that for the publishers that do the utmost to provide readers with a high-quality experience, such latency is capable of significantly degrading the user experience.
Interestingly, we find that the top 500 (in Alexa ranking) websites exhibited significantly lower \hb latency than the rest of websites.

Although \hblong tech promises multiple, in-parallel bid requests to \DPs that can provide the best possible ad price to the publisher, Javascript on the users' end is single-threaded.
This means that even if the \hb provider's wrapper performs well-optimized asynchronous ad calls, these still need to stand in the network queue, thus increasing not only the overall \hb execution time but also the entire webpage's loading time.
These delays can have adverse effects on user's browsing experience while loading a \hb-enabled webpage.
Interestingly, we find that the 10 most popular \DPs exhibit lower variability in latency than the rest, demonstrating that they invest appropriate resources to reduce latencies at the user-end.

\subsection{Late Bids: Revenue \& Network Cost}

The broadcasting nature of \hblong results in an enormous amount of bid requests to multiple \DPs.
As measured in this study, a typical website has a small number of available ad-slots (\ie 2-6 ad-slots for the median case, depending on the type of \hb), but some auctions request for more ad-slots than they have available to show (even up to 20).
Side banner and top banner are the most popular ad-slots auctioned in \hb.
For each ad-slot, a parallel auction takes place that requests bids from numerous DSPs.
As expected, the more ad-slots present in a webpage, the higher the overall \hb latency: when 1-3 ad-slots are auctioned, the median latency is 0.3-0.57 seconds, but for 3-5 slots, the median latency is 0.57-0.92 seconds.

This overwhelming volume of bid requests significantly increases the needed processing power for ADXs and the decision engines of DSPs, thus skyrocketing their infrastructure costs~\cite{infra-cost}.
Indeed, companies that started supporting \hb experienced increases of up to 100\% in the bid requests they received~\cite{infra-cost2} (\ie between 5 million and 6 million requests per second) for the very same number of available ad-slots as before.
Interestingly, the same partners may in fact compete for the same ad-slots more than once: in the \hb, and then in the regular waterfall model, since the publisher may still fall back to the waterfall if the \hb does not reach high enough prices for the auctioned slots~\cite{infra-cost}.

Apart from skyrocketing the infrastructure costs, the increased amount of bid requests also increases the response time for DSPs, causing lots of delayed bids.
We found that in more than 50\% of auctions, half of bid responses arrive too late (after the publisher's set threshold) to be considered, due to high latency.
These late bids not only are wasted network resources and processing power from the point of view of the \DPs, but also loss of potentially higher revenues for publishers.

\subsection{Limitations \& Future Work}

The present work is a first, comprehensive study of the \hb protocol, and an effort to measure the ecosystem and partners involved.
Unfortunately, \hb documentation is scarce at best, and it has been a feat to reverse-engineer the protocol, and understand the numerous \hb libraries used by the crawled websites.
Due to several limitations in the data collection process, in the present study, we focused on specific dimensions and left others for future work.
In a follow-up work on \hb, it is important to address the limitations of our \toolname, and also perform extensions to study aspects not covered in this work:
\begin{itemize} [leftmargin=0.5cm]
\item Perform extensive analysis of all available \hb libraries, to increase coverage of the ecosystem and websites employing \hb.
\item Study in detail the impact that \hb has on the UX and page load time of each website, as well as
the various hosting infrastructures responsible for the crawled websites, locations of \DPs involved, and categories of websites to find associations with \hb prices and latencies.
\item Investigate the privacy of online users accessing \hb-enabled websites for potential PII leaks, and measure the impact that \hb may have on user anonymity, as well as the use of HTTP vs. HTTPs for \hb transactions.
\end{itemize}

\section*{Acknowledgements}
We thank the anonymous reviewers and our shepherd Narseo Vallina Rodriguez, for the insightful comments and help to improve the present manuscript.
The research leading to these results has received funding from EU Marie Sklodowska-Curie and Horizon 2020 Research \& Innovation Programme under grant agreements 690972 and 786669, respectively.
The paper reflects only the authors' view and the Agency and the Commission are not responsible for any use that may be made of the information it contains.

\bibliographystyle{plain}
\balance
\bibliography{main}
\end{document}